\documentclass{article}
\usepackage{graphicx}  % standard LaTeX graphics tool;
\usepackage{amsmath}   % For getting proper math eqns
\usepackage[compress]{cite}
\usepackage{amssymb}   % Used for getting various symbols eg. gtrsim, lesssim etc.
\usepackage{bm} % For bold math (esp of lower greek letters)
\usepackage{dcolumn}% Align table columns on decimal point
\usepackage{color}
\usepackage{mathrsfs}
\usepackage{amsfonts}
\usepackage{varioref}
\usepackage{textcomp}
\usepackage[margin=1in]{geometry}
\RequirePackage[colorlinks,citecolor=blue,urlcolor=magenta,linkcolor=blue]{hyperref}
\allowdisplaybreaks
\def\EH{Einstein-Hilbert }

\def\gr{general relativity}
\def\RN{Reissner-Nordstr\"{o}m }

\def\tln{tidal Love number}
\labelformat{section}{Section #1} 
\labelformat{subsection}{Section #1} 
\labelformat{subsubsection}{Section #1}
\labelformat{subsubsubsection}{Section #1}
\labelformat{equation}{Eq.~(#1)} 
\labelformat{figure}{Fig.~#1} 
\labelformat{subfigure}{Fig.~\thefigure#1} 
\labelformat{table}{Tab.~#1} 
\labelformat{appendix}{Appendix #1}
\title{Tidal Love Numbers of Black Holes and Neutron Stars in the Presence of Higher Dimensions: Implications of GW170817}
\author{Kabir Chakravarti\footnote{kabir@iucaa.in}~$^{1}$, Sumanta Chakraborty\footnote{sumantac.physics@gmail.com}~$^{2}$, Sukanta Bose\footnote{sukanta@iucaa.in}~$^{1,3}$ and Soumitra SenGupta\footnote{tpssg@iacs.res.in}~$^{2}$
\\
{$^{1}$\small{IUCAA, Post Bag 4, Ganeshkhind, pune University Campus, Pune 411007, India}}\\
{$^{2}$\small{School of Physical Sciences, Indian Association for the Cultivation of Science, Kolkata-700032, India}}\\
{$^{3}$\small{Department of Physics and Astronomy, Washington State University}}\\
{\small{1245 Webster, Pullman, WA 99164-2814, USA}}}
\begin{document}
  
\maketitle
%%%%%%%%%%%%%%%%%%%%%%%%%%%%%%%%%%%%%%%%%%%%%%%%%%%%%%%%%%%%%%%%%%%%%%%%%%%%%%%%%%%%%%%%%%%%%%%%%%%
%%%%%%%%%%%%%%%%%%%%%%%%%%%%%%%%%%%%%%%%%%%%%%%%%%%%%%%%%%%%%%%%%%%%%%%%%%%%%%%%%%%%%%%%%%%%%%%%%%%
%%%%%%%%%%%%%%%%%%%%%%%%%%%%%%%%%%%%%%%%%%%%%%%%%%%%%%%%%%%%%%%%%%%%%%%%%%%%%%%%%%%%%%%%%%%%%%%%%%%
\begin{abstract}
We calculate the tidal Love numbers of black holes and neutron stars in the presence of higher dimensions. The perturbation equations around an arbitrary static and spherically symmetric metric for the even parity modes are presented in the context of an \emph{effective} four-dimensional theory on the brane. This subsequently leads to the sought expression for the tidal Love number for black holes in the presence of extra spatial dimensions. Surprisingly, these numbers are non-zero and, more importantly, negative. We extend our method to determine the tidal Love number of neutron stars in a spacetime inheriting extra dimensions and show that, in the context of effective gravitational theory on the brane, they are smaller than in general relativity. Finally we have explicitly demonstrated that earlier constraints on the parameters inherited from higher dimensions are consistent with the bound on the tidal deformability parameter from the GW170817 event as well.

\end{abstract}
%%%%%%%%%%%%%%%%%%%%%%%%%%%%%%%%%%%%%%%%%%%%%%%%%%%%%%%%%%%%%%%%%%%%%%%%%%%%%%%%%%%%%%%%%%%%%%%%%%%
%%%%%%%%%%%%%%%%%%%%%%%%%%%%%%%%%%%%%%%%%%%%%%%%%%%%%%%%%%%%%%%%%%%%%%%%%%%%%%%%%%%%%%%%%%%%%%%%%%%
%%%%%%%%%%%%%%%%%%%%%%%%%%%%%%%%%%%%%%%%%%%%%%%%%%%%%%%%%%%%%%%%%%%%%%%%%%%%%%%%%%%%%%%%%%%%%%%%%%%
%\newpage
%%%%%%%%%%%%%%%%%%%%%%%%%%%%%%%%%%%%%%%%%%%%%%%%%%%%%%%%%%%%%%%%%%%%%%%%%%%%%%%%%%%%%%%%%%%%%%%%%%%
%%%%%%%%%%%%%%%%%%%%%%%%%%%%%%%%%%%%%%%%%%%%%%%%%%%%%%%%%%%%%%%%%%%%%%%%%%%%%%%%%%%%%%%%%%%%%%%%%%%
%%%%%%%%%%%%%%%%%%%%%%%%%%%%%%%%%%%%%%%%%%%%%%%%%%%%%%%%%%%%%%%%%%%%%%%%%%%%%%%%%%%%%%%%%%%%%%%%%%%
\section{Introduction and Motivation}\label{TLN_v1_Introduction}

The recent success of Gravitational Wave (GW) detectors in detecting GW from mergers of binary Black Holes (BHs) and Neutron Stars (NSs) has been a feather in the cap of scientific and technological advancement \cite{Abbott:2017vtc,TheLIGOScientific:2016pea,Abbott:2016nmj,TheLIGOScientific:2016src,Abbott:2016blz,TheLIGOScientific:2017qsa,Abbott:2018exr}. As the GW detectors get more and more sensitive, it will enable us to make more and more specific statements with regard to different astronomical aspects of GWs. Thus with GW astronomy established on a firm footing, we can now turn our attention to answer some of the more fundamental questions of nature. These include: (a) subtle relations regarding black hole physics (e.g., no-hair theorem, area increase theorem, existence of black hole horizons themselves) \cite{Loeb:2013lfa,Cabero:2017avf,Cardoso:2016rao,Abedi:2016hgu,Wade:2013hoa,Cardoso:2017cqb}; (b) evidence of theories beyond general relativity, in particular, presence of higher curvature terms in the \EH action, theories involving specific curvature scalar coupling, known as Horndeski theories are of significant interest among others \cite{Berti:2015itd,Babichev:2016jom,Bhattacharya:2016naa,Creminelli:2017sry,Kase:2018aps,Bhattacharyya:2017tyc,Pratten:2015rqa,Mukherjee:2017fqz,Banerjee:2017hzw,Sakstein:2016oel,Sakstein:2017xjx,Casalino:2018wnc} and finally (c) fundamental structure of the spacetime itself \cite{Seahra:2004fg,Clarkson:2005eb,Seahra:2005us,Seahra:2009gw,Witek:2013ora,Chakraborty:2017qve,Andriot:2017oaz}.  An interesting line of thought in understanding the fundamental structure of spacetime happens to be the question of presence (or absence) of extra spatial dimensions to our usual four-dimensional spacetime. 

Incidentally, the idea of existence of spatial extra dimensions originated for a completely different reason. Originally, these extra dimensions have appeared quite naturally in string theory, whose existence requires ten or more dimensions. However later, the existence of extra dimensions returned to the community in order to solve the `gauge hierarchy problem' \cite{ArkaniHamed:1998rs,Antoniadis:1998ig,Antoniadis:1990ew,Rubakov:1983bz,Rubakov:1983bb,PerezLorenzana:2005iv,Csaki:2004ay}. This has to do with the fact that the scale of physics appears very much un-correlated and hierarchical in nature. This is because the scale of electro-weak symmetry breaking, which is $\sim 10^{3}~\textrm{GeV}$, appears to be completely disconnected from the Planck scale at $\sim 10^{18}~\textrm{GeV}$. Another and probably more practical reason to worry about the gauge hierarchy problem has to do with the running of the mass of the Higgs Boson. In order to get the observed value of the Higg's mass at the Large Hadron Collider (LHC), one needs to fine tune the counter term arising out of renormalization to one part in $10^{15}$. This large fine tuning is another reason for the existence of gauge hierarchy problem \cite{PerezLorenzana:2005iv,Sundrum:2005jf,Csaki:2004ay}. 

In the context of extra dimensions it is indeed possible to cure the gauge hierarchy problem. Broadly speaking, there are two possibilities to achieve the same. First, one can introduce large extra dimensions such that even though the four dimensional Planck's constant is large enough, the fundamental higher dimensional Planck's constant is in the TeV scale due to volume suppression \cite{ArkaniHamed:1998rs,Antoniadis:1998ig,Antoniadis:1990ew,PerezLorenzana:2005iv}. The above model does not incorporate gravitational dynamics and hence not of much significance from the GW point of view. On the other hand, it is also possible to provide a gravitational resolution to the gauge hierarchy problem. One of the many possible ways to achieve the same is by introducing warped Anti-de Sitter geometry in the bulk, resulting into exponential suppression of physical scales on the brane \cite{Randall:1999ee,Goldberger:1999uk,Randall:1999vf,Davoudiasl:1999tf}. We will mainly work with the second possibility, where gravity itself comes to the rescue. 

However, since all the observations regarding GWs are being made in the four dimensional spacetime it is legitimate to ask for an effective four dimensional theory starting from the original higher dimensional one \cite{Randall:1999ee,Garriga:1999yh,Chamblin:1999by,Davoudiasl:2000wi,Dadhich:2000am,Maartens:2001jx,Gregory:2008rf,Emparan:1999wa,Gregory:2011kh,Lehner:2011wc,Frolov:2009jr}. As such we have modelled the extra dimension in a simple manner, having one additional spacelike dimension. By imposing orbifold symmetry on the extra dimension one can determine the effective gravitational field equation inheriting additional corrections from the bulk geometry \cite{Shiromizu:1999wj,Dadhich:2000am,Maartens:2001jx,Maartens:2003tw}. Such effective gravitational field equations have been derived in the context of general relativity in \cite{Shiromizu:1999wj}, which subsequently was generalized to many other situations as well \cite{Chakraborty:2014xla,Chakraborty:2015taq,Chakraborty:2015bja}. In this work we will content ourselves with the general relativistic situation alone and with the advent of GW astronomy, we would like to put the theory to the strong field test of gravitational interaction. 

The presence of extra dimensions would leave very specific signatures on the GWs emitted by merging BHs or NSs. In principle, these signatures will be present for the entire duration of the GW signal, often separated into three distinct regimes, commonly known in GW terminology as the inspiral, merger and ringdown phase. The inspiral and merger phase would require a detailed numerical analysis, which is presently unavailable. While the ringdown phase can be understood analytically, since it involves computation of Quasi Normal Modes (QNMs) from perturbed BHs in these theories \cite{Chandrasekhar:1985kt,Nollert:1999ji,Kokkotas:1999bd,Berti:2009kk,Konoplya:2011qq,Konoplya:2008yy,Konoplya:2006br,Kanti:2005xa}. For both the higher dimensional black holes or black holes in the effective gravitational theory one can indeed obtain distinct signatures of extra dimensions in the QNMs, which recently have been investigated in some detail in the literature \cite{Toshmatov:2016bsb,Chakraborty:2017qve} (also see \cite{Aneesh:2018hlp,Visinelli:2017bny}). 

In this paper, we have set out to explore yet another observational window put forward by the recent GW observations, namely the modifications to the tidal Love number due to the presence of higher dimensions (for some recent works regarding \tln\ in other theories beyond \gr, see \cite{Cardoso:2017cfl,Yazadjiev:2018xxk}). The \tln\ for a neutron star or black hole essentially corresponds to the deformation in the respective objects caused by an external tidal field \cite{Lattimer:2004pg,Hinderer:2007mb,Flanagan:2007ix,Binnington:2009bb,Damour:2009vw,Hinderer:2009ca,DelPozzo:2013ala,Poisson:2014gka,Porto:2016zng,Yagi:2016ejg,Barausse:2018vdb,Chakravarti:2018uyi}. Interestingly, for black holes in \gr\ the \tln\ identically vanishes and hence determining the \tln\ for black holes in theories beyond \gr\ is of significant importance. If the associated \tln\ for black holes are non-zero, then it may provide a crucial hint towards these theories beyond \gr\ \cite{Cardoso:2017cfl}. On the other hand, the \tln\ for neutron stars are also of sufficient importance in the context of equation of state (EoS) of the material forming the neutron star. Since the composition of the neutron stars are not very well known, there does not exist much tight constraints on the EoS parameter either. Given the recent detection of NS-NS merger in the Advanced LIGO detectors, it is desirable to understand the EoS parameter of NSs either by appropriate theoretical modelling or using numerical simulations. The best way to get an understanding of the \tln\ is from the early regime of the in-spiral phase, as the signal is very clean. However, the influence of tidal effects is through the phase of the waveform and is only a small correction. Thus to detect the same one may invoke the matched filtering technique by integrating the measured waveform, such that the accumulated phase shift due to the tidal corrections becomes larger. The influence of the internal structure of the neutron star on the gravitational wave phase is characterized by the ratio of the induced quadrupole moment to the perturbing external tidal field and is denoted by $\lambda$. This is related to the dimensionless \tln\ $k_{2}$ through the following relation: $(3/2)(G_{4}\lambda/R^{5})$, where $G_{4}$ is the four dimensional gravitational constant and $R$ is the radius of the neutron star. A systematic study of the determination of the tidal Love number from perturbations of Einstein?s gravitational field equations was presented in Refs. \cite{Hinderer:2007mb,Flanagan:2007ix} and later used extensively in the GW literature, where the ideas were both refined and broadened \cite{Binnington:2009bb,Damour:2009vw}. In this paper, we would like to understand the modifications brought in the \tln\ of both black holes and neutron stars due to the presence of extra dimensions. This is evidently the first step in trying to understand the complete set of modifications to the GW signal which may arise in the strong field regime of gravity.

The paper is organized as follows: In \ref{TLN_v1_Perturbation} we discuss the general framework of gravitational perturbation outside a compact object, which could be either a black hole or a neutron star. The formalism derived here has been subsequently applied in \ref{TLN_v1_Black} to discuss the \tln\ for a black hole on the four dimensional spacetime in presence of higher dimensions. Finally we also demonstrate the modifications to the \tln\ associated with neutron stars pertaining to the existence of higher dimensions. Implications of the results derived here for current and future GW merger events have been explored in \ref{imp_GW170817}. We finally conclude with a discussion on the results obtained. Some relevant computations have also been presented in \ref{TLN_App_v2_Diff}.

\textit{Notations and Conventions:} Throughout the paper we have assumed $\hbar=1=c$. Greek indices are used to present the four dimensional quantities and we work with the mostly positive signature convention. By and large, in this paper by the phrase \tln\ we will essentially mean the dimensionfull \tln\ $\lambda$. Whenever the dimensionless \tln\ will be used it will be mentioned explicitly.
%%%%%%%%%%%%%%%%%%%%%%%%%%%%%%%%%%%%%%%%%%%%%%%%%%%%%%%%%%%%%%%%%%%%%%%%%%%%%%%%%%%%%%%%%%%%%%%%%%%
%%%%%%%%%%%%%%%%%%%%%%%%%%%%%%%%%%%%%%%%%%%%%%%%%%%%%%%%%%%%%%%%%%%%%%%%%%%%%%%%%%%%%%%%%%%%%%%%%%%
%%%%%%%%%%%%%%%%%%%%%%%%%%%%%%%%%%%%%%%%%%%%%%%%%%%%%%%%%%%%%%%%%%%%%%%%%%%%%%%%%%%%%%%%%%%%%%%%%%%
\section{Perturbation outside a compact object in presence of extra dimension: General analysis}\label{TLN_v1_Perturbation}

In this section, we will determine the master equation satisfied by the even parity gravitational perturbations, essential for computation of electric type tidal Love number, in the exterior region of a compact object, which could be either a black hole or a neutron star. In the case of a black hole, the exterior equation alone is sufficient to determine the \tln. This will help us to determine whether the electric type tidal Love number can have non-zero value in a black hole background, in the presence of extra dimensions. On the other hand, for neutron stars, unlike the case for black holes, understanding the gravitational perturbation in the exterior is not sufficient to determine the \tln, one needs to solve the gravitational perturbation in the interior of the star as well, which we will take up in the subsequent sections. Since the structure of the gravitational field equations in the exterior is of importance, irrespective of the nature of the compact object, we will first characterize the background spacetime around which perturbations will be considered in some detail, before deriving the associated master equation for gravitational perturbation around this background with full generality.
%%%%%%%%%%%%%%%%%%%%%%%%%%%%%%%%%%%%%%%%%%%%%%%%%%%%%%%%%%%%%%%%%%%%%
%%%%%%%%%%%%%%%%%%%%%%%%%%%%%%%%%%%%%%%%%%%%%%%%%%%%%%%%%%%%%%%%%%%%%
%%%%%%%%%%%%%%%%%%%%%%%%%%%%%%%%%%%%%%%%%%%%%%%%%%%%%%%%%%%%%%%%%%%%%
\subsection{Setting up the background spacetime}\label{TLN_v1_Background}

The spacetime under consideration inherits one additional spatial dimension and we are interested in its effect on the gravitational field equations associated with four dimensional brane hypersurface we are living in. This can be achieved by starting from the five dimensional bulk Einstein's equations and then projecting on the lower dimensional brane hypersurface, using appropriate normal vectors. Since there is no matter field present in the exterior region, the sole contribution will come from gravitational effects. In such vacuum exterior spacetime, the projected field equation describing the dynamics of gravity can be written as \cite{Maartens:2003tw},
%%%%%%%%%%%%%%%%%%%%%%%%%%%%%%%%%%%%%%%%%%%%%%%%%%%%%%%%%%%%%%
\begin{align}\label{TLN_v1_01}
~^{(4)}G_{\mu \nu}+E_{\mu \nu}=0~.
\end{align}
%%%%%%%%%%%%%%%%%%%%%%%%%%%%%%%%%%%%%%%%%%%%%%%%%%%%%%%%%%%%%%
Here $E_{\mu \nu}$ corresponds to the electric part of the bulk Weyl tensor and $^{(4)}G_{\mu \nu}$ is the Einstein tensor projected on the brane starting from the bulk. Interestingly the derivation of the above equation uses minimal information about the nature of the extra dimension, in particular it only requires the brane to be located at a orbifold fixed point associated with the extra dimension. Thus the result presented in this work will be general and possibly will hold for a large class of extra dimensions. This fact is reflected in the determination of $E_{\mu \nu}$, requiring information about the bulk spacetime, which in general is not available. Thus following \cite{Dadhich:2000am,Maartens:2001jx,Harko:2004ui,Maartens:2003tw} we will assume that $E_{\mu \nu}$ can be represented by a perfect fluid, defined by energy density $U$ and pressure $P$. The exact correspondence of the structure of $E_{\mu \nu}$ with that of bulk spacetime can be understood as and when we get some handle on the nature of physical theories near Planck scale. 

Since by definition the computation of \tln\ involves an equilibrium configuration, we will consider a static and spherically symmetric spacetime as the one depicting the background geometry. The line element for such a spacetime can always be casted as:
%%%%%%%%%%%%%%%%%%%%%%%%%%%%%%%%%%%%%%%%%%%%%%%%%%%%%%%%%%%%%%
\begin{align}\label{TLN_v1_02}
ds^{2}=-e^{2\nu(r)}dt^{2}+e^{2\lambda(r)}dr^{2}+r^{2}d\Omega ^{2}~,
\end{align}
%%%%%%%%%%%%%%%%%%%%%%%%%%%%%%%%%%%%%%%%%%%%%%%%%%%%%%%%%%%%%%
where $\nu(r)$ and $\lambda(r)$ are arbitrary functions of the radial coordinate alone. Therefore, the associated field equations simplify to
%%%%%%%%%%%%%%%%%%%%%%%%%%%%%%%%%%%%%%%%%%%%%%%%%%%%%%%%%%%%%%
\begin{align}
e^{-2\lambda(r)}\left(\frac{1}{r^{2}}-\frac{2\lambda'}{r}\right)-\frac{1}{r^{2}}&=-24\pi \tilde{U}(r)~,
\label{TLN_v1_03a}
\\
e^{-2\lambda(r)}\left(\frac{2\nu'}{r}+\frac{1}{r^{2}}\right)-\frac{1}{r^{2}}&=8\pi \left(\tilde{U}+2\tilde{P}\right)~,
\label{TLN_v1_03b}
\\
e^{-2\lambda}\left\{\nu''+\nu'^{2}-\nu'\lambda'+\frac{1}{r}\left(\nu'-\lambda'\right)\right\}&=8\pi\left(\tilde{U}-\tilde{P}\right)~.
\label{TLN_v1_03c}
\end{align}
%%%%%%%%%%%%%%%%%%%%%%%%%%%%%%%%%%%%%%%%%%%%%%%%%%%%%%%%%%%%%%
Here we have defined
%%%%%%%%%%%%%%%%%%%%%%%%%%%%%%%%%%%%%%%%%%%%%%%%%%%%%%%%%%%%%%
\begin{align}\label{TLN_v1_04}
\tilde{U}=\frac{2G_{4}}{\left(8\pi G_{4}\right)^{2}\lambda _{\rm b}}U;\qquad \tilde{P}=\frac{2G_{4}}{\left(8\pi G_{4}\right)^{2}\lambda _{\rm b}}P~,
\end{align}
%%%%%%%%%%%%%%%%%%%%%%%%%%%%%%%%%%%%%%%%%%%%%%%%%%%%%%%%%%%%%%
where $G_{4}$ is the four-dimensional Newton's constant and $\lambda _{\rm b}$ is the brane tension \cite{Dadhich:2000am,Maartens:2001jx,Maartens:2003tw}. As evident from the above equations, the quantities $\tilde{U}$ and $\tilde{P}$ (or, $U$ and $P$) encodes all the higher dimensional effects and hence are completely determined by $E_{\mu \nu}$. In particular, the ``dark radiation" term $U$ is related to $E_{\mu \nu}$ as $U=-(G_{4}/G_{5})^{2}E_{\mu \nu}u^{\mu}u^{\nu}$, where $G_{5}$ is the five-dimensional gravitational constant. While the ``dark pressure" term $P$ essentially originates from the spatially trace-free and symmetric part of $E_{\mu \nu}$. One can also verify the correctness of the above equations from dimensional analysis as well. Thus the above set of equations can be thought of as the background equations in presence of an anisotropic fluid, such that: $\rho=3\tilde{U}$, $p_{r}=\tilde{U}+2\tilde{P}$ and $p_{\perp}=\tilde{U}-\tilde{P}$. Given this structure, we would like to determine the master equation governing the even parity gravitational perturbation, which we elaborate in the next section. 
%%%%%%%%%%%%%%%%%%%%%%%%%%%%%%%%%%%%%%%%%%%%%%%%%%%%%%%%%%%%%%%%%%%%%
%%%%%%%%%%%%%%%%%%%%%%%%%%%%%%%%%%%%%%%%%%%%%%%%%%%%%%%%%%%%%%%%%%%%%
%%%%%%%%%%%%%%%%%%%%%%%%%%%%%%%%%%%%%%%%%%%%%%%%%%%%%%%%%%%%%%%%%%%%%
\subsection{The equations governing even parity perturbations}\label{TLN_v1_Even}

Having described the background gravitational field equations, depending on the metric functions $\nu(r)$ and $\mu(r)$ along with the effect of extra dimensions (encoded in $\tilde{U}$ and $\tilde{P}$), let us concentrate on the structure of the perturbed equations. These are obtained from the first order variation of the background metric presented in \ref{TLN_v1_01}; i.e., one considers a modified metric $g_{\mu \nu}^{\rm mod}=g_{\mu \nu}+h_{\mu \nu}$ and hence computes the differential equations that 
$h_{\mu \nu}$ satisfies to its leading order. The perturbation $h_{\mu \nu}$ can further be subdivided into two classes, namely even and odd, depending on its transformation under parity. The electric type tidal Love numbers are governed by the even parity modes, while the magnetic type tidal Love numbers are dictated by the odd parity modes. Since the even parity modes have direct observational consequences to GW physics, in this work we will concentrate on the electric type tidal Love numbers with the hope of returning to the discussion of magnetic type tidal Love numbers elsewhere. Thus we will exclusively work with even parity gravitational perturbations. These even parity perturbations are characterized by three functions, $H_{0}(r)$, $H_{2}(r)$ and $K(r)$, as follows: 
%%%%%%%%%%%%%%%%%%%%%%%%%%%%%%%%%%%%%%%%%%%%%%%%%%%%%%%%%%%%%%
\begin{align}\label{TLN_v3_01}
h_{\mu \nu;\ell m}^{\rm even}=\textrm{diag}\left[e^{-2\nu(r)}H_{0},e^{2\lambda(r)}H_{2}(r),r^{2}K(r),r^{2}\sin^{2}\theta K(r) \right]Y_{\ell m}(\theta,\phi)~.
\end{align}
%%%%%%%%%%%%%%%%%%%%%%%%%%%%%%%%%%%%%%%%%%%%%%%%%%%%%%%%%%%%%%
Here we are working with a fixed value of $\ell$ and $m$, which will suffice as the determination of \tln\ requires $\ell=2$ but is independent of the choice of $m$. For the sake of generality, we will keep $\ell$ arbitrary for the moment, but will set $\ell=2$ as and when necessary.

In general, the ten components of the gravitational field equations will lead to ten perturbed equations. On the other hand, except for the $(r,\theta)$ component, all the other off-diagonal entries in the perturbed equations are trivially satisfied.  Among others, the right hand side of the angular components of the perturbation equations are identical, hence $\delta G^{\theta}_{\theta}-\delta G^{\phi}_{\phi}=0$, which yields $H_{2}=H_{0}\equiv H(r)$. Moreover, $\delta G^{r}_{\theta}=8\pi G_{4} \delta T^{r}_{\theta}$ results in $K'=H'+2\nu'H$. Thus, one need not consider both $H_{0}$ and $H_{2}$ as two independent perturbation components; rather, one can concentrate on the differential equation satisfied by $H$ alone. This will also help to determine the nature of the remaining perturbation component $K(r)$ through the relation $K'=H'+2\nu'H$ introduced earlier. To proceed further, we need to take care of the addition of the angular components as well as the temporal and radial components of the perturbation equation. 

First, the addition of the angular components, i.e., the equation $\delta G^{\theta}_{\theta}+\delta G^{\phi}_{\phi}=8\pi (\delta T^{\theta}_{\theta}+\delta T^{\phi}_{\phi})$ yields,
%%%%%%%%%%%%%%%%%%%%%%%%%%%%%%%%%%%%%%%%%%%%%%%%%%%%%%%%%%%%%%
\begin{align}\label{TLN_v1_05}
e^{-2\lambda}r^{2}K''+e^{-2\lambda}rK'\left\{2+r\left(\nu'-\lambda'\right)\right\}-e^{-2\lambda}r^{2}H''
&-e^{-2\lambda}rH'\left(3r\nu'-r\lambda'+2\right)
\nonumber
\\
&-16\pi r^{2}\delta p_{\perp}-16\pi r^{2}Hp_{\perp}=0~.
\end{align}
%%%%%%%%%%%%%%%%%%%%%%%%%%%%%%%%%%%%%%%%%%%%%%%%%%%%%%%%%%%%%%
Here, $p_{\perp}=\tilde{U}-\tilde{P}$, as defined earlier. Thus, the above equation depends on double derivatives of both $H(r)$ and $K(r)$. Proceeding further, we can use the radial component of the perturbation equation, which reads $\delta G^{r}_{r}=8\pi \delta T^{r}_{r}$. Expressing in terms of derivatives of the perturbed and unperturbed metric components, we obtain,
%%%%%%%%%%%%%%%%%%%%%%%%%%%%%%%%%%%%%%%%%%%%%%%%%%%%%%%%%%%%%%
\begin{align}\label{TLN_v1_06}
e^{-2\lambda}\left(1+r\nu'\right)rK'-\left\{\frac{1}{2}\ell\left(\ell+1\right)-1\right\}K-e^{-2\lambda}rH'
+\left\{\frac{1}{2}\ell\left(\ell+1\right)-1-8\pi r^{2}p_{r}\right\}H-8\pi r^{2}\delta p_{r}=0~.
\end{align}
%%%%%%%%%%%%%%%%%%%%%%%%%%%%%%%%%%%%%%%%%%%%%%%%%%%%%%%%%%%%%%
Here the radial pressure $p_{r}$ is defined as $\tilde{U}+2\tilde{P}$, and inherits the contributions from the existence of higher dimensions. Finally, the remaining perturbation equation $\delta G^{t}_{t}=8\pi \delta T^{t}_{t}$ can be expressed in terms of metric perturbations as,
%%%%%%%%%%%%%%%%%%%%%%%%%%%%%%%%%%%%%%%%%%%%%%%%%%%%%%%%%%%%%%
\begin{align}\label{TLN_v1_07}
e^{-2\lambda}r^{2}K''+e^{-2\lambda}rK'\left(3-r\lambda'\right)
&-\left\{\frac{1}{2}\ell (\ell+1)-1\right\}K-re^{-2\lambda}H'
\nonumber
\\
&-\left\{\frac{1}{2}\ell\left(\ell+1\right)+1-8\pi r^{2}\rho\right\}H+8\pi r^{2}\delta \rho =0~.
\end{align}
%%%%%%%%%%%%%%%%%%%%%%%%%%%%%%%%%%%%%%%%%%%%%%%%%%%%%%%%%%%%%%
The density $\rho$ defined above, just as $p_{r}$ and $p_{\perp}$, is expressible in terms of $\tilde{U}$ and $\tilde{P}$, such that $\rho \equiv 3\tilde{U}$.

Having written down all the equations governing the perturbations, the aim is to arrive at another equation that does not involve any perturbation in the matter sector. In other words, we want to eliminate $\delta \tilde{U}$ and $\delta \tilde{P}$ from the above equations. As a first step in that direction, we consider the subtraction of the $(t,t)$ and the $(r,r)$ components of the perturbation equation, i.e., subtraction of \ref{TLN_v1_06} from \ref{TLN_v1_07}, which yields,
%%%%%%%%%%%%%%%%%%%%%%%%%%%%%%%%%%%%%%%%%%%%%%%%%%%%%%%%%%%%%%
\begin{align}\label{TLN_v1_09}
e^{-2\lambda}r^{2}K''+e^{-2\lambda}rK'\left\{2-r\left(\lambda'+\nu'\right)\right\}&-\ell\left(\ell+1\right)H
\nonumber
\\
&+16\pi r^{2}\left(2\tilde{U}+\tilde{P}\right)H+16\pi r^{2}\left\{1+2\left(\frac{\partial \tilde{U}}{\partial \tilde{P}}\right)\right\}\delta \tilde{P}=0~.
\end{align}
%%%%%%%%%%%%%%%%%%%%%%%%%%%%%%%%%%%%%%%%%%%%%%%%%%%%%%%%%%%%%%
In the above expression we have replaced $\rho$ and $p_{r}$ in terms of $\tilde{U}$ and $\tilde{P}$, and used $\delta \tilde{U}=(\partial \tilde{U}/\partial \tilde{P})\delta \tilde{P}$. In an identical fashion, we can rewrite \ref{TLN_v1_05} also in terms of $\tilde{U}$ and $\tilde{P}$, which will involve a $\delta \tilde{P}$ term. Thus one can use it to eliminate the $\delta \tilde{P} $ term appearing in \ref{TLN_v1_09}, which results in
%%%%%%%%%%%%%%%%%%%%%%%%%%%%%%%%%%%%%%%%%%%%%%%%%%%%%%%%%%%%%%
\begin{align}\label{TLN_v1_11}
\Bigg(\frac{\partial \tilde{U}}{\partial \tilde{P}}&-1\Bigg)e^{-2\lambda}r^{2}K''
+\left(\frac{\partial \tilde{U}}{\partial \tilde{P}}-1\right)e^{-2\lambda}rK'\left\{2-r\left(\lambda'+\nu'\right)\right\}
-\left(\frac{\partial \tilde{U}}{\partial \tilde{P}}-1\right)\ell\left(\ell+1\right)H
\nonumber
\\
&+16\pi r^{2}\left(2\tilde{U}+\tilde{P}\right)H\left(\frac{\partial \tilde{U}}{\partial \tilde{P}}-1\right)
+\left\{1+2\left(\frac{\partial \tilde{U}}{\partial \tilde{P}}\right)\right\}
\Bigg[e^{-2\lambda}r^{2}K''-e^{-2\lambda}r^{2}H''
\nonumber
\\
&\qquad+e^{-2\lambda}\left\{2+\left(\nu'-\lambda'\right)\right\}rK'
-e^{-2\lambda}\left(3r\nu'-r\lambda'+2\right)rH'-16\pi r^{2}H\left(\tilde{U}-\tilde{P}\right)\Bigg]=0~.
\end{align}
%%%%%%%%%%%%%%%%%%%%%%%%%%%%%%%%%%%%%%%%%%%%%%%%%%%%%%%%%%%%%%
As desired, the above equation involves no reference to the $\delta \tilde{U}$ or $\delta \tilde{P}$ term whatsoever. Even though the above equation depends on both $H$ and $K$, a close inspection reveals that it depends on $K$ only through its derivative. Thus one may use the result, $K'=H'+2\nu'H$, to transform the above differential equation to one that depends solely on $H(r)$. This results in the following differential equation for $H(r)$,
%%%%%%%%%%%%%%%%%%%%%%%%%%%%%%%%%%%%%%%%%%%%%%%%%%%%%%%%%%%%%%
\begin{align}\label{TLN_v1_13}
&\Bigg(3\frac{\partial \tilde{U}}{\partial \tilde{P}}\Bigg)e^{-2\lambda}r^{2}\left(H''+2\nu'H'+2\nu''H\right)
+e^{-2\lambda}r\left(H'+2\nu'H\right)\left\{\left(2-r\lambda'\right)\left(3\frac{\partial \tilde{U}}{\partial \tilde{P}}\right)+r\nu'\left(2+\frac{\partial \tilde{U}}{\partial \tilde{P}}\right)\right\}
\nonumber
\\
&-e^{-2\lambda}\left(3r\nu'-r\lambda'+2\right)rH'\left\{1+2\left(\frac{\partial \tilde{U}}{\partial \tilde{P}}\right)\right\}
+16\pi r^{2}\left(2\tilde{U}+\tilde{P}\right)H\left(\frac{\partial \tilde{U}}{\partial \tilde{P}}-1\right)
\nonumber
\\
&-16\pi r^{2}H\left(\tilde{U}-\tilde{P}\right)\left\{1+2\left(\frac{\partial \tilde{U}}{\partial \tilde{P}}\right)\right\}
-\left\{1+2\left(\frac{\partial \tilde{U}}{\partial \tilde{P}}\right)\right\}e^{-2\lambda}r^{2}H''
-\left(\frac{\partial \tilde{U}}{\partial \tilde{P}}-1\right)\ell\left(\ell+1\right)H=0~.
\end{align}
%%%%%%%%%%%%%%%%%%%%%%%%%%%%%%%%%%%%%%%%%%%%%%%%%%%%%%%%%%%%%%
This explicitly depicts how one may manipulate all the perturbation equations so as to eliminate any term depending on $\delta \tilde{P}$ and $\delta \tilde{U}$ along with any remaining gravitational perturbation components in order to arrive at a single equation for the gravitational perturbation $H$. In the next section, we will manipulate these terms to express the above equation in a more tractable form, which can subsequently be used to determine the \tln. 
%%%%%%%%%%%%%%%%%%%%%%%%%%%%%%%%%%%%%%%%%%%%%%%%%%%%%%%%%%%%%%%%%%%%%
%%%%%%%%%%%%%%%%%%%%%%%%%%%%%%%%%%%%%%%%%%%%%%%%%%%%%%%%%%%%%%%%%%%%%
%%%%%%%%%%%%%%%%%%%%%%%%%%%%%%%%%%%%%%%%%%%%%%%%%%%%%%%%%%%%%%%%%%%%%
\subsection{The master equation outside a compact object in presence of higher dimensions}\label{TLN_v1_Master}

In this section, we will determine the master equation satisfied by the even parity gravitational perturbation, introduced in \ref{TLN_v3_01}, outside a compact object in presence of an extra dimension. For that purpose, one may start by computing the coefficients of $H''$, $H'$ and $H$ in the equation for single gravitational perturbation $H(r)$ presented in \ref{TLN_v1_13}. It turns out that all these coefficients simplify by a great amount, ultimately resulting into the following structure
%%%%%%%%%%%%%%%%%%%%%%%%%%%%%%%%%%%%%%%%%%%%%%%%%%%%%%%%%%%%%%
\begin{align}\label{TLN_v1_15}
e^{-2\lambda}&r^{2}\left(\frac{\partial \tilde{U}}{\partial \tilde{P}}-1\right)H''+re^{-2\lambda}\left\{\frac{\partial \tilde{U}}{\partial \tilde{P}}-1\right\}\left(2-r\lambda'+r\nu'\right)H'
+e^{-2\lambda}r^{2}\left(6\frac{\partial \tilde{U}}{\partial \tilde{P}}\right)\left(\nu''-\nu'\lambda'+\nu'^{2}\right)H
\nonumber
\\
&-4\nu'^{2}e^{-2\lambda}r^{2}\left(\frac{\partial \tilde{U}}{\partial \tilde{P}}\right)H
+12re^{-\lambda}\nu'\left(\frac{\partial \tilde{U}}{\partial \tilde{P}}\right)H+4\nu'^{2}e^{-2\lambda}r^{2}H
-\left(\frac{\partial \tilde{U}}{\partial \tilde{P}}-1\right)\ell\left(\ell+1\right)H
\nonumber
\\
&+16\pi r^{2}\left(2\tilde{U}+\tilde{P}\right)H\left(\frac{\partial \tilde{U}}{\partial \tilde{P}}-1\right)
-16\pi r^{2}H\left(\tilde{U}-\tilde{P}\right)\left\{1+2\left(\frac{\partial \tilde{U}}{\partial \tilde{P}}\right)\right\}=0~.
\end{align}
%%%%%%%%%%%%%%%%%%%%%%%%%%%%%%%%%%%%%%%%%%%%%%%%%%%%%%%%%%%%%%
The coefficients of all the derivatives of $H$ can be easily read off from the above expression. In particular, the structural similarity between the coefficients of $H''$ and $H'$ suggests to divide the above equation throughout by $r^{2}e^{-2\lambda}\{(\partial \tilde{U}/\partial \tilde{P})-1\}$, which yields,
%%%%%%%%%%%%%%%%%%%%%%%%%%%%%%%%%%%%%%%%%%%%%%%%%%%%%%%%%%%%%%
\begin{align}\label{TLN_v1_17}
H''&+\left(\frac{2}{r}-\lambda'+\nu'\right)H'+6\frac{(\partial \tilde{U}/\partial \tilde{P})}{(\partial \tilde{U}/\partial \tilde{P})-1}\left(\nu''-\nu'\lambda'+\nu'^{2}+\frac{2\nu'}{r}\right)H-4\nu'^{2}H
\nonumber
\\
&-e^{2\lambda}\frac{\ell\left(\ell+1\right)}{r^{2}}H
+16\pi e^{2\lambda}\left(2\tilde{U}+\tilde{P}\right)H
-16\pi e^{2\lambda}H\left(\tilde{U}-\tilde{P}\right)\frac{\left\{1+2(\partial \tilde{U}/\partial \tilde{P})\right\}}{(\partial \tilde{U}/\partial \tilde{P})-1}=0~.
\end{align}
%%%%%%%%%%%%%%%%%%%%%%%%%%%%%%%%%%%%%%%%%%%%%%%%%%%%%%%%%%%%%%
In order to arrive at the above expression, we have manipulated various terms appearing in the coefficient of $H(r)$. The above expression can be written in a more suggestive form, if we keep in mind that so far we have not used the background field equations. In particular, we can use the background field equations to replace $\nu(r)$ and $\lambda(r)$ by a more suitable expression. First of all, one can integrate \ref{TLN_v1_03a} in order to yield,
%%%%%%%%%%%%%%%%%%%%%%%%%%%%%%%%%%%%%%%%%%%%%%%%%%%%%%%%%%%%%%
\begin{align}\label{TLN_v1_19}
e^{-2\lambda}=1-\frac{2\tilde{m}(r)}{r};\qquad \tilde{m}(r)\equiv G_{4}M+12\pi \int dr~ r^{2}\tilde{U}(r)~,
\end{align}
%%%%%%%%%%%%%%%%%%%%%%%%%%%%%%%%%%%%%%%%%%%%%%%%%%%%%%%%%%%%%%
where $M$ is the mass of the central compact object. One can further use \ref{TLN_v1_03a} as well as \ref{TLN_v1_03b} in order to arrive at 
%%%%%%%%%%%%%%%%%%%%%%%%%%%%%%%%%%%%%%%%%%%%%%%%%%%%%%%%%%%%%%
\begin{align}\label{TLN_v1_20}
r\left(\nu'-\lambda'\right)=8\pi r^{2}e^{2\lambda}\left(\tilde{P}-\tilde{U}\right)+\frac{2\tilde{m}(r)}{r}e^{2\lambda}~,
\end{align}
%%%%%%%%%%%%%%%%%%%%%%%%%%%%%%%%%%%%%%%%%%%%%%%%%%%%%%%%%%%%%%
where \ref{TLN_v1_19} has also been used. Another significant relation can be derived by using \ref{TLN_v1_03c}:
%%%%%%%%%%%%%%%%%%%%%%%%%%%%%%%%%%%%%%%%%%%%%%%%%%%%%%%%%%%%%%
\begin{align}
\nu''+\nu'^{2}-\nu'\lambda'&=8\pi e^{2\lambda}\left(\tilde{U}-\tilde{P}\right)-\frac{1}{r}\left(\nu'-\lambda'\right)~.
\label{TLN_v1_18c}
\end{align}
%%%%%%%%%%%%%%%%%%%%%%%%%%%%%%%%%%%%%%%%%%%%%%%%%%%%%%%%%%%%%%
Note that the last term can again be written in terms of $\tilde{U}$ and $\tilde{P}$ by using \ref{TLN_v1_20}. Thus in \ref{TLN_v1_17} we can use both \ref{TLN_v1_19} and \ref{TLN_v1_18c} to write down all the background quantities in terms of $\tilde{U}$, $\tilde{P}$ and the derivative $(\partial \tilde{U}/\partial \tilde{P})$. This results in the following structure of the master equation for even parity gravitational perturbation,
%%%%%%%%%%%%%%%%%%%%%%%%%%%%%%%%%%%%%%%%%%%%%%%%%%%%%%%%%%%%%%
\begin{align}\label{TLN_v1_23}
H''&+\left\{\frac{2}{r}+8\pi re^{2\lambda}\left(\tilde{P}-\tilde{U}\right)+\frac{2\tilde{m}(r)}{r^{2}}e^{2\lambda}\right\}H'+6\frac{(\partial \tilde{U}/\partial \tilde{P})}{(\partial \tilde{U}/\partial \tilde{P})-1}
\left\{24\pi e^{2\lambda}\tilde{U}\right\}H
\nonumber
\\
&\left\{-4\nu'^{2}-e^{2\lambda}\frac{\ell\left(\ell+1\right)}{r^{2}}
+16\pi e^{2\lambda}\left(2\tilde{U}+\tilde{P}\right)\right\}H
-16\pi e^{2\lambda}H\left(\tilde{U}-\tilde{P}\right)\frac{\left\{1+2(\partial \tilde{U}/\partial \tilde{P})\right\}}{(\partial \tilde{U}/\partial \tilde{P})-1}=0~.
\end{align}
%%%%%%%%%%%%%%%%%%%%%%%%%%%%%%%%%%%%%%%%%%%%%%%%%%%%%%%%%%%%%%
As is evident, the only information about background spacetime is through the $e^{2\lambda}$ and $\nu'^{2}$ terms; the rest of the terms have been converted to some combinations of the dark radiation and dark pressure terms, which carry imprints of the presence of higher dimensions. The above expression can further be simplified by grouping various terms appropriately, appearing in the coefficient of $H(r)$. In particular, by expressing $(\partial \tilde{U}/\partial \tilde{P})=\{(\partial \tilde{U}/\partial \tilde{P})-1\}+1$ we can write down the final compact expression for the differential equation satisfied by the perturbation $H(r)$ as,
%%%%%%%%%%%%%%%%%%%%%%%%%%%%%%%%%%%%%%%%%%%%%%%%%%%%%%%%%%%%%%
\begin{align}\label{TLN_v1_26}
H''+\left\{\frac{2}{r}+8\pi re^{2\lambda}\left(\tilde{P}-\tilde{U}\right)+\frac{2\tilde{m}(r)}{r^{2}}e^{2\lambda}\right\}H'
&+\left\{-4\nu'^{2}-e^{2\lambda}\frac{\ell\left(\ell+1\right)}{r^{2}}
+16\pi e^{2\lambda}\left(9\tilde{U}+3\tilde{P}\right)\right\}H
\nonumber
\\
&+\frac{16\pi e^{2\lambda}\left(6\tilde{U}+3\tilde{P}\right)}{(\partial \tilde{U}/\partial \tilde{P})-1}H=0~
\end{align}
%%%%%%%%%%%%%%%%%%%%%%%%%%%%%%%%%%%%%%%%%%%%%%%%%%%%%%%%%%%%%%
which is the result we were after. Note that this is a single differential equation for $H(r)$, one of the perturbation components of the even parity metric perturbation, and hence depicts the master equation that one must solve. An understanding of $H(r)$ will also lead to comprehending the other metric perturbation components. However for arbitrary choices of $\tilde{U}$ and $\tilde{P}$ this is as far as we can go; to proceed further one needs to have a relation between $\tilde{U}$ and $\tilde{P}$. Only then can one explicitly compute the solution to the above equation, either analytically or by numerical techniques. 
%%%%%%%%%%%%%%%%%%%%%%%%%%%%%%%%%%%%%%%%%%%%%%%%%%%%%%%%%%%%%%%%%%%%%%%%%%%%%%%%%%%%%%%%%%%%%%%%%%%
%%%%%%%%%%%%%%%%%%%%%%%%%%%%%%%%%%%%%%%%%%%%%%%%%%%%%%%%%%%%%%%%%%%%%%%%%%%%%%%%%%%%%%%%%%%%%%%%%%%
%%%%%%%%%%%%%%%%%%%%%%%%%%%%%%%%%%%%%%%%%%%%%%%%%%%%%%%%%%%%%%%%%%%%%%%%%%%%%%%%%%%%%%%%%%%%%%%%%%%
\section{Computation of tidal Love number for braneworld black hole}\label{TLN_v1_Black}

We have already developed the basic equation governing even parity  gravitational perturbations in the exterior region of a compact object. If the compact object depicts a neutron star, we have to write down the corresponding equations in its interior as well before obtaining the associated \tln. However for black holes the situation is much simpler, we just have to solve the exterior solution and use some suitable boundary conditions requiring regularity at the black hole horizon. Keeping this in mind, in this section we will discuss the tidal Love number for braneworld black holes. As mentioned earlier, this requires some choices for the dark radiation term $\tilde{U}$ and dark pressure term $\tilde{P}$, for which we will use the most favoured equation of state for the Weyl fluid (given by $E_{\mu \nu}$). This will help us to explicitly write down the background solution and extract information about the equation of state parameter. That in turn will help us present \ref{TLN_v1_26} in a more appropriate form satisfied by the gravitational perturbation $H(r)$, which we will solve subsequently to find the electric type \tln.  
%%%%%%%%%%%%%%%%%%%%%%%%%%%%%%%%%%%%%%%%%%%%%%%%%%%%%%%%%%%%%%%%%%%%%
%%%%%%%%%%%%%%%%%%%%%%%%%%%%%%%%%%%%%%%%%%%%%%%%%%%%%%%%%%%%%%%%%%%%%
%%%%%%%%%%%%%%%%%%%%%%%%%%%%%%%%%%%%%%%%%%%%%%%%%%%%%%%%%%%%%%%%%%%%%
\subsection{The background spacetime}\label{TLN_v1_Particular}

In this section we will present the background spacetime, given an appropriate equation of state parameter, depicting a black hole solution on the brane \cite{Dadhich:2000am}. As evident from \ref{TLN_v1_26}, in the context of Weyl fluid induced from higher dimensional spacetime, the most interesting equation of state parameter, which simplifies the background and perturbation equations a lot corresponds to, $2\tilde{U}+\tilde{P}=0$ \cite{Dadhich:2000am,Harko:2004ui}. Thus we have the equation of state parameter to be, $(\partial \tilde{U}/\partial \tilde{P})=-(1/2)$. With this choice, the differential equation for the background spacetime, presented in \ref{TLN_v1_03a}-\ref{TLN_v1_03c}, can be explicitly solved with the following structure for the Weyl fluid,
%%%%%%%%%%%%%%%%%%%%%%%%%%%%%%%%%%%%%%%%%%%%%%%%%%%%%%%%%%%%%%
\begin{align}\label{TLN_v1_27}
\tilde{U}=-\frac{\tilde{P}_{0}}{2r^{4}};\qquad \tilde{P}=\frac{\tilde{P}_{0}}{r^{4}}~,
\end{align}
%%%%%%%%%%%%%%%%%%%%%%%%%%%%%%%%%%%%%%%%%%%%%%%%%%%%%%%%%%%%%%
where $\tilde{P}_{0}$ is a constant dependent on the nature of the bulk spacetime, i.e., it provides the signature of the existence of higher dimensions. The metric elements, on the other hand, take the following form,
%%%%%%%%%%%%%%%%%%%%%%%%%%%%%%%%%%%%%%%%%%%%%%%%%%%%%%%%%%%%%%
\begin{align}\label{TLN_v1_28}
e^{2\nu}=e^{-2\lambda}=1-\frac{2G_{4}M}{r}-\frac{12\pi \tilde{P}_{0}}{r^{2}}~.
\end{align}
%%%%%%%%%%%%%%%%%%%%%%%%%%%%%%%%%%%%%%%%%%%%%%%%%%%%%%%%%%%%%%
This is alike the \RN black hole, with one crucial difference. In the context of \RN black hole coefficient of $r^{-2}$ must be positive, while in the present context it can be positive or negative depending on the sign of $\tilde{P}_{0}$.  In what follows we will assume $\tilde{P}_{0}>0$, which will be the only non-trivial possibility when we consider the case of the exterior of a compact object in the next section. Keeping that in mind, in the context of black hole as well we will consider the case in which the coefficient of $r^{-2}$ term is negative. Given the above structure of the metric elements associated with the background spacetime, it immediately follows that,
%%%%%%%%%%%%%%%%%%%%%%%%%%%%%%%%%%%%%%%%%%%%%%%%%%%%%%%%%%%%%%
\begin{align}\label{TLN_v1_29}
2\nu'=\left(\frac{2G_{4}M}{r^{2}}+\frac{24\pi \tilde{P}_{0}}{r^{3}}\right)\left(1-\frac{2G_{4}M}{r}-\frac{12\pi \tilde{P}_{0}}{r^{2}}\right)^{-1}=-2\lambda'~,
\end{align}
%%%%%%%%%%%%%%%%%%%%%%%%%%%%%%%%%%%%%%%%%%%%%%%%%%%%%%%%%%%%%%
as well as in this particular case we have, $\tilde{m}(r)=G_{4}M+(6\pi \tilde{P}_{0}/r)$. These expressions will be used extensively in the later sections, when we will explicitly compute the \tln\ for the above black hole solution located on the brane. 

Before we delve into the computation of \tln, let us briefly discuss how to even define the \tln\ in the present context. For that purpose, suppose that the above system is placed in a static, external quadrupolar field $\mathcal{E}_{ij}$,, where $i$ and $j$ are spatial indices. In response to the above, the compact object will develop a quadrupole moment $Q_{ij}$ and in the asymptotic rest frame the temporal components of the metric element, in the present context, can be written as \cite{Hinderer:2007mb},
%%%%%%%%%%%%%%%%%%%%%%%%%%%%%%%%%%%%%%%%%%%%%%%%%%%%%%%%%%%%%%
\begin{align}\label{TLN_v2_N01}
\frac{1}{2}\left(1+g_{tt}\right)=\frac{G_{4}M}{r}+\frac{\beta}{2}\left(\frac{G_{4}M}{r}\right)^{2}
+\frac{3G_{4}Q_{ij}}{2r^{3}}\left(n^{i}n^{j}-\frac{1}{3}\delta ^{ij}\right)+\mathcal{O}\left(\frac{1}{r^{4}}\right)
-\frac{1}{2}\mathcal{E}_{ij}n^{i}n^{j}r^{2}+\mathcal{O}\left(r^{3}\right)~,
\end{align}
%%%%%%%%%%%%%%%%%%%%%%%%%%%%%%%%%%%%%%%%%%%%%%%%%%%%%%%%%%%%%%
where we have introduced a dimensionless constant $\beta \equiv (12\pi \tilde{P}_{0}/G_{4}^{2}M^{2})$, which has been inherited from extra dimension. As evident in the $\beta \rightarrow 0$ limit we recover the expansion of $g_{tt}$ as in the context of \gr. Further, the quantity $n^{i}\equiv x^{i}/r$ and the above expansion defines the quadrupole moment $Q_{ij}$ of the compact object  and the external quadrupolar field $\mathcal{E}_{ij}$. The proportionality constant between them corresponds to the \tln\ $\lambda$, defined as: $Q_{ij}=-\lambda \mathcal{E}_{ij}$. Further we can expand both $Q_{ij}$ and $\mathcal{E}_{ij}$ in terms of spherical harmonics, such that
%%%%%%%%%%%%%%%%%%%%%%%%%%%%%%%%%%%%%%%%%%%%%%%%%%%%%%%%%%%%%%
\begin{align}\label{TLN_v2_N03}
\mathcal{E}_{ij}=\sum _{m=-2}^{2}\mathcal{E}_{m}Y_{(2m)~ij};\qquad Q_{ij}=\sum _{m=-2}^{2}Q_{m}Y_{(2m)~ij}~,
\end{align}
%%%%%%%%%%%%%%%%%%%%%%%%%%%%%%%%%%%%%%%%%%%%%%%%%%%%%%%%%%%%%%
where $Y_{(2m)~ij}$ are traceless tensors related to the $\ell=2$ spherical harmonics $Y_{2m}(\theta,\phi)$ by the following relation: $Y_{2m}=Y_{(2m)~ij}n^{i}n^{j}$. Thus using the expressions for $\mathcal{E}_{ij}$ and $Q_{ij}$ presented in \ref{TLN_v2_N03}, the expansion of the temporal component of the metric becomes,
%%%%%%%%%%%%%%%%%%%%%%%%%%%%%%%%%%%%%%%%%%%%%%%%%%%%%%%%%%%%%%
\begin{align}\label{TLN_v2_N04}
\frac{1}{2}\left(1+g_{tt}\right)=\frac{G_{4}M}{r}+\frac{\beta}{2}\left(\frac{G_{4}M}{r}\right)^{2}
+\frac{3G_{4}}{2r^{3}}\sum _{m=-2}^{2}Q_{m}Y_{2m}+\mathcal{O}\left(\frac{1}{r^{4}}\right)
-\frac{1}{2}\sum _{m=-2}^{2}\mathcal{E}_{m}Y_{2m}r^{2}+\mathcal{O}\left(r^{3}\right)~.
\end{align}
%%%%%%%%%%%%%%%%%%%%%%%%%%%%%%%%%%%%%%%%%%%%%%%%%%%%%%%%%%%%%%
Since the \tln\ does not depend on the specific value of $m$, it still can be determined from the relation $Q_{m}=-\lambda \mathcal{E}_{m}$. Thus the \tln\ is derived as follows: (a) using the background spacetime geometry presented in \ref{TLN_v1_27} and \ref{TLN_v1_28} one can rewrite the perturbation equation presented in \ref{TLN_v1_26}; (b) this can be solved subsequently to determine $H(r)$ and, hence, $h_{00}=e^{-2\nu}H_{0}(r)$. Then one can use the expansion of the temporal component of the metric element, presented in \ref{TLN_v2_N04}, to determine $Q_{m}$ and $\mathcal{E}_{m}$, which in turn determines the \tln. In the next section, we will demonstrate the final form of the perturbation equation in the present context. 
%%%%%%%%%%%%%%%%%%%%%%%%%%%%%%%%%%%%%%%%%%%%%%%%%%%%%%%%%%%%%%%%%%%%%
%%%%%%%%%%%%%%%%%%%%%%%%%%%%%%%%%%%%%%%%%%%%%%%%%%%%%%%%%%%%%%%%%%%%%
%%%%%%%%%%%%%%%%%%%%%%%%%%%%%%%%%%%%%%%%%%%%%%%%%%%%%%%%%%%%%%%%%%%%% 
\subsection{Derivation of the final form of the master equation}\label{TLN_v1_Derivation}

Having spelt out the structure of the background spacetime along with the notion of tidal Love number, let us concentrate on the derivation of the final form of the master equation. For this we rewrite \ref{TLN_v1_26} using the background metric elements presented in \ref{TLN_v1_27} and \ref{TLN_v1_28}. This yields the following differential equation for $H(r)$ in the background of a braneworld black hole,
%%%%%%%%%%%%%%%%%%%%%%%%%%%%%%%%%%%%%%%%%%%%%%%%%%%%%%%%%%%%%%
\begin{align}\label{TLN_v1_31}
\Big(1&-\frac{2G_{4}M}{r}-\frac{12\pi \tilde{P}_{0}}{r^{2}}\Big)H''+\left\{\frac{2}{r}\left(1-\frac{2G_{4}M}{r}-\frac{12\pi \tilde{P}_{0}}{r^{2}}\right)+8\pi r\left(\frac{3\tilde{P}_{0}}{2r^{4}}\right)+\frac{2}{r^{2}}\left(G_{4}M+\frac{6\pi \tilde{P}_{0}}{r} \right)\right\}H'
\nonumber
\\
&+\Bigg\{-\left(\frac{2G_{4}M}{r^{2}}+\frac{24\pi \tilde{P}_{0}}{r^{3}}\right)^{2}\left(1-\frac{2G_{4}M}{r}-\frac{12\pi \tilde{P}_{0}}{r^{2}}\right)^{-1}-\frac{\ell\left(\ell+1\right)}{r^{2}}+16\pi \left(-\frac{3\tilde{P}_{0}}{2r^{4}}\right)\Bigg\}H=0~,
\end{align}
%%%%%%%%%%%%%%%%%%%%%%%%%%%%%%%%%%%%%%%%%%%%%%%%%%%%%%%%%%%%%%
where both sides of the original equation have been multiplied by $e^{-2\lambda}$. It is always advantageous to rewrite any equation in terms of dimensionless quantities, which prompts us to introduce a new dimensionless variable $x$ and replace the radial coordinate $r$, such that $x=(r/G_{4}M)-1$. This results in the transformation of $H'$ and $H''$, such that $H'$ gets scaled by $(G_{4}M)^{-1}$, while $H''$ is modified by $(G_{4}M)^{-2}$. Thus, finally multiplying \ref{TLN_v1_31} throughout by $r^{2}$ and subsequently introducing the variable $x$ in appropriate places, while removing the radial coordinate we obtain,
%%%%%%%%%%%%%%%%%%%%%%%%%%%%%%%%%%%%%%%%%%%%%%%%%%%%%%%%%%%%%%
\begin{align}\label{TLN_v1_36}
\left\{x^{2}-1-\beta\right\}\partial _{x}^{2}H+2x\partial _{x}H
+\left\{-\ell\left(\ell+1\right)-\frac{4\left(x+1+\beta\right)^{2}}{(1+x)^{2}\left(x^{2}-1-\beta\right)}
-\frac{2\beta}{(1+x)^{2}}\right\}H=0~.
\end{align}
%%%%%%%%%%%%%%%%%%%%%%%%%%%%%%%%%%%%%%%%%%%%%%%%%%%%%%%%%%%%%%
The above simple structure of the gravitational perturbation equation is obtained by again introducing the dimensionless quantity $\beta \equiv (12\pi \tilde{P}_{0}/G_{4}^{2}M^{2})$. Note that for $\beta=0$, i.e., when extra dimensional effects are absent the above equation reduces to,
%%%%%%%%%%%%%%%%%%%%%%%%%%%%%%%%%%%%%%%%%%%%%%%%%%%%%%%%%%%%%%
\begin{align}\label{TLN_v1_37}
\left\{x^{2}-1\right\}\partial _{x}^{2}H+2x\partial _{x}H
+\left\{-\ell\left(\ell+1\right)-\frac{4}{\left(x^{2}-1\right)}\right\}H=0~,
\end{align}
%%%%%%%%%%%%%%%%%%%%%%%%%%%%%%%%%%%%%%%%%%%%%%%%%%%%%%%%%%%%%%
exactly coinciding with the result derived in \cite{Hinderer:2007mb} in the context of \gr. This acts as an acid test of the formalism developed above since it explicitly demonstrates the correctness of our result by reproduces the general relativistic result in an appropriate limit. We will now discuss the solution of the above equation in the asymptotic limit in order to determine the associated \tln. For this purpose we will be using \ref{TLN_v2_N04}, the asymptotic expansion of the temporal component of the perturbed metric. 
%%%%%%%%%%%%%%%%%%%%%%%%%%%%%%%%%%%%%%%%%%%%%%%%%%%%%%%%%%%%%%%%%%%%%
%%%%%%%%%%%%%%%%%%%%%%%%%%%%%%%%%%%%%%%%%%%%%%%%%%%%%%%%%%%%%%%%%%%%%
%%%%%%%%%%%%%%%%%%%%%%%%%%%%%%%%%%%%%%%%%%%%%%%%%%%%%%%%%%%%%%%%%%%%%
\subsection{Tidal Love numbers of braneworld black holes}\label{TLN_v1_LoveBlack}

In this section we will explicitly compute the tidal Love number for the above black hole solution in the presence of extra dimension. As an aside, we will also demonstrate why the \tln\ for general relativistic black holes must vanish, while they can be non-zero in the present context\footnote{Of course, the fact that \tln\ for black holes in \gr\ must vanish, holds when the expansion of temporal component of metric element is truncated as presented in \ref{TLN_v2_N04}. If higher order corrections are taken into account, the \tln\ for black holes in \gr\ may turn out to be non-zero.}. However, unlike the case of \gr, where an exact solution to the perturbation equation presented in \ref{TLN_v1_37} is possible in terms of Legendre polynomials, in present context a general analytic solution seems difficult. Before commenting on the possibility of getting an analytic solution, let us write down \ref{TLN_v1_36} in a more suggestive form. This can be achieved by introducing a new variable $y$, related to the old one as: $x=\sqrt{1+\beta}~y$. Thus the terms involving $\partial _{x}H$ and $\partial _{x}^{2}H$ will get modified by introduction of $(1+\beta)^{-1/2}$ and $(1+\beta)^{-1}$ respectively, such that the master equation for $H(x)$ will now be converted into a master equation for $H(y)$, which reads,
%%%%%%%%%%%%%%%%%%%%%%%%%%%%%%%%%%%%%%%%%%%%%%%%%%%%%%%%%%%%%%
\begin{align}\label{TLN_v2_39}
\left\{y^{2}-1\right\}\partial _{y}^{2}H+2y\partial _{y}H
-\left\{\ell\left(\ell+1\right)+\frac{4}{\left(y^{2}-1\right)}\frac{\left(y+\sqrt{1+\beta}\right)^{2}}{(\sqrt{1+\beta}~y+1)^{2}}
+\frac{2\beta}{(\sqrt{1+\beta}~y+1)^{2}}\right\}H=0~.
\end{align}
%%%%%%%%%%%%%%%%%%%%%%%%%%%%%%%%%%%%%%%%%%%%%%%%%%%%%%%%%%%%%%
As evident, in terms of the new variable $y$, the coefficient of $\partial _{y}^{2}H$ and $\partial _{y}H$ are identical to the corresponding differential equation for \gr, however the term proportional to $H(y)$ differs significantly. Due to the complicated nature of the coefficient of $H(y)$, this differential equation, unlike the $\beta=0$ case, does not have any general analytic solution. But in order to determine the \tln\ it is sufficient that we understand the asymptotic limit and as we will demonstrate below, an analytic solution can indeed be obtained. Since, $y=(1+\beta)^{-1/2}\{(r/M)-1\}$, the asymptotic, i.e., large $r$ limit implies $y\rightarrow \infty$. Hence the above differential equation simplifies significantly and we obtain,
%%%%%%%%%%%%%%%%%%%%%%%%%%%%%%%%%%%%%%%%%%%%%%%%%%%%%%%%%%%%%%
\begin{align}\label{TLN_v2_41}
\left\{y^{2}-1\right\}\partial _{y}^{2}H+2y\partial _{y}H
-\left\{\ell\left(\ell+1\right)+\frac{4}{y^{2}-1}\left(\frac{1+\frac{\beta}{2}}{1+\beta}\right)\right\}H=0~.
\end{align}
%%%%%%%%%%%%%%%%%%%%%%%%%%%%%%%%%%%%%%%%%%%%%%%%%%%%%%%%%%%%%%
The fact that general relativistic result is reproduced in the $\beta\rightarrow 0$ limit is apparent from the above differential equation. It turns out that the above equation admits analytic solutions in terms of associated Legendre polynomials. The details of the solution and asymptotic limits of the associated Legendre polynomial has been presented in \ref{TLN_App_v2_Diff}, which interested reader may refer to. But for our purpose, we can take a cue from \ref{TLN_App_v2_Diff} (see in particular, \ref{TLN_v1_App_20} presented in the appendix) to write down the asymptotic solution of the above differential equation with $\ell=2$ as,
%%%%%%%%%%%%%%%%%%%%%%%%%%%%%%%%%%%%%%%%%%%%%%%%%%%%%%%%%%%%%%
\begin{align}\label{TLN_v2_42}
H(y)=\left\lbrace \frac{3A_{1}\sqrt{\pi}}{\Gamma(3-\bar{\beta})} \right\rbrace y^{2}
+\left\lbrace -\frac{A_{1}}{15\Gamma(-2-\bar{\beta})} 
+\frac{B_{1}\Gamma(3+\bar{\beta}) e^{i\pi \bar{\beta}}}{15} \right\rbrace y^{-3}
\end{align}
%%%%%%%%%%%%%%%%%%%%%%%%%%%%%%%%%%%%%%%%%%%%%%%%%%%%%%%%%%%%%%
where, we have defined $\bar{\beta}^{2}=4\{1+(\beta/2)\}\{1+\beta \}^{-1}$ and $A_{1}$, $B_{1}$ are arbitrary constants of integration. At this stage in order to compare with the asymptotic expansion of the metric elements it is essential to re-introduce the radial co-ordinate $r$ through the following relation: $y=x(1+\beta)^{-1/2}$, with $x=(r/G_{4}M)-1$. In the asymptotic limit, we obtain the following structure for metric perturbation,
%%%%%%%%%%%%%%%%%%%%%%%%%%%%%%%%%%%%%%%%%%%%%%%%%%%%%%%%%%%%%%
\begin{align}\label{TLN_v2_43}
H(\beta;r)=\left\lbrace \frac{3A_{1}\sqrt{\pi}}{\Gamma(3-\bar{\beta})\left(1+\beta\right)} \right\rbrace 
&\times \left(\frac{r}{G_{4}M}\right)^{2}
\nonumber
\\
&+\left(1+\beta\right)^{3/2}\left\lbrace -\frac{A_{1}}{15\Gamma(-2-\bar{\beta})} 
+\frac{B_{1}\Gamma(3+\bar{\beta}) e^{i\pi \bar{\beta}}}{15} \right\rbrace \left(\frac{G_{4}M}{r}\right)^{3}
\end{align}
%%%%%%%%%%%%%%%%%%%%%%%%%%%%%%%%%%%%%%%%%%%%%%%%%%%%%%%%%%%%%%
To ensure the correctness of the above result, we must demonstrate that it is consistent with \gr, which follows from considering the $\beta \rightarrow 0$ limit of \ref{TLN_v2_43}, which yields,
%%%%%%%%%%%%%%%%%%%%%%%%%%%%%%%%%%%%%%%%%%%%%%%%%%%%%%%%%%%%%%
\begin{align}\label{TLN_v2_44}
H_{\rm GR}(r)\equiv H(\beta=0;r)=\left\lbrace \frac{3A_{1}\sqrt{\pi}}{\Gamma(1)}\right\rbrace 
\left(\frac{r}{G_{4}M}\right)^{2}
+\left\lbrace \frac{8B_{1}}{5} \right\rbrace \left(\frac{G_{4}M}{r}\right)^{3}
\end{align}
%%%%%%%%%%%%%%%%%%%%%%%%%%%%%%%%%%%%%%%%%%%%%%%%%%%%%%%%%%%%%%
where we have used the fact that $\bar{\beta}\rightarrow 2$ as the parameter $\beta$ vanishes. Further, since by definition $\Gamma (1)=1$, it follows from the identity, $\Gamma(1)=0\times \Gamma(0)$, that $\Gamma(0)$ diverges. Thus given that $\Gamma(-4)=(-1/4)(-1/3)(-1/2)(-1)\Gamma(0)$, we can immediately argue that $\Gamma(-4)$ diverges. The fact that $\Gamma(-4)$ diverges has been used in order to arrive at \ref{TLN_v2_44}. As is evident, the solution for the metric perturbation, presented in \ref{TLN_v2_44}, matches exactly with the general-relativistic result presented in \cite{Hinderer:2007mb}, provided we introduce a new constant $A=A_{1}\sqrt{\pi}$. This suggests to use the same constant in the general case as well, besides we also introduce $B=B_{1}\exp(i\pi \mu)$. Thus in terms of the newly defined arbitrary constants $A$ and $B$, the solution to the metric perturbation in the background of braneworld black hole takes the following structure,
%%%%%%%%%%%%%%%%%%%%%%%%%%%%%%%%%%%%%%%%%%%%%%%%%%%%%%%%%%%%%%
\begin{align}\label{TLN_v2_46}
H(r)=\left\lbrace \frac{3A}{\Gamma(3-\bar{\beta})\left(1+\beta\right)} \right\rbrace 
&\times \left(\frac{r}{G_{4}M}\right)^{2}
\nonumber
\\
&+\left(1+\beta\right)^{3/2}\left\lbrace -\frac{A}{15\sqrt{\pi}\Gamma(-2-\bar{\beta})} 
+\frac{B\Gamma(3+\bar{\beta})}{15} \right\rbrace \left(\frac{G_{4}M}{r}\right)^{3}~,
\end{align}
%%%%%%%%%%%%%%%%%%%%%%%%%%%%%%%%%%%%%%%%%%%%%%%%%%%%%%%%%%%%%%
which has the correct general relativistic limit. Thus at both the differential-equation and asymptotic-solution levels, we have explicitly verified that the general relativistic result can be reproduced in the appropriate limit. Now that we know the exact form for the asymptotic solution, we can compute the \tln. This will essentially follow from \ref{TLN_v2_N01}. In presence of perturbation, we have the temporal component of the metric involving perturbation to be $g^{\rm mod}_{tt}=-e^{2\nu}-e^{-\nu}H(r)Y_{2m}$, where $e^{2\nu}$ is given by \ref{TLN_v1_28} and $H(r)$ by \ref{TLN_v2_46}. Hence using the expressions for $e^{2\nu}$ and $H(r)$, we obtain the following relation in the asymptotic limit,
%%%%%%%%%%%%%%%%%%%%%%%%%%%%%%%%%%%%%%%%%%%%%%%%%%%%%%%%%%%%%%
\begin{align}\label{TLN_v2_N05}
\frac{G_{4}M}{r}&+\frac{\beta}{2}\left(\frac{G_{4}M}{r}\right)^{2}
+\frac{3G_{4}}{2r^{3}}\sum _{m=-2}^{2}Q_{m}Y_{2m}+\mathcal{O}\left(\frac{1}{r^{3}}\right)
-\frac{1}{2}\sum _{m=-2}^{2}\mathcal{E}_{m}Y_{2m}r^{2}+\mathcal{O}\left(r^{3}\right)
\nonumber
\\
&=\frac{G_{4}M}{r}+\frac{\beta}{2}\left(\frac{G_{4}M}{r}\right)^{2}
-\frac{1}{2}\sum _{m=-2}^{2}\left\lbrace \frac{3A}{\Gamma(3-\bar{\beta})\left(1+\beta\right)} \right\rbrace 
\left(\frac{r}{G_{4}M}\right)^{2}Y_{2m}
\nonumber
\\
&-\frac{1}{2}\left(1+\beta\right)^{3/2}\sum _{m=-2}^{2}\left\lbrace -\frac{A}{15\sqrt{\pi}\Gamma(-2-\bar{\beta})} 
+\frac{B\Gamma(3+\bar{\beta})}{15} \right\rbrace \left(\frac{r}{G_{4}M}\right)^{-3}Y_{2m}~.
\end{align}
%%%%%%%%%%%%%%%%%%%%%%%%%%%%%%%%%%%%%%%%%%%%%%%%%%%%%%%%%%%%%%
As it turns out, the unperturbed components exactly cancels out and subsequent matching of the powers of $r$ on both sides of the above equation yield, 
%%%%%%%%%%%%%%%%%%%%%%%%%%%%%%%%%%%%%%%%%%%%%%%%%%%%%%%%%%%%%%
\begin{align}\label{TLN_v2_N06}
\left\lbrace \frac{3A}{\Gamma(3-\bar{\beta})\left(1+\beta\right)} \right\rbrace 
=\left(G_{4}M\right)^{2}\mathcal{E}_{m};
\quad 
\left(1+\beta\right)^{3/2}\left\lbrace -\frac{A}{15\sqrt{\pi}\Gamma(-2-\bar{\beta})} 
+\frac{B\Gamma(3+\bar{\beta})}{15} \right\rbrace
=-3\frac{G_{4}Q_{m}}{\left(G_{4}M\right)^{3}}
\end{align}
%%%%%%%%%%%%%%%%%%%%%%%%%%%%%%%%%%%%%%%%%%%%%%%%%%%%%%%%%%%%%%
Given the above relation, the \tln\ can be easily determined by first taking the ratio of the above equations, and then using the definition, $Q_{m}=-\lambda \mathcal{E}_{m}$. Performing the above we obtain the tidal Love number to be,
%%%%%%%%%%%%%%%%%%%%%%%%%%%%%%%%%%%%%%%%%%%%%%%%%%%%%%%%%%%%%%
\begin{align}\label{TLN_v2_N07}
G_{4}\lambda\equiv \frac{1}{3}\left(1+\beta\right)^{5/2}\left\lbrace -\frac{\Gamma(3-\bar{\beta})}{45\sqrt{\pi}\Gamma(-2-\bar{\beta})} 
+\frac{B\Gamma(3+\bar{\beta})\Gamma(3-\bar{\beta})}{45A} \right\rbrace \left(G_{4}M\right)^{5}~,
\end{align}
%%%%%%%%%%%%%%%%%%%%%%%%%%%%%%%%%%%%%%%%%%%%%%%%%%%%%%%%%%%%%%
where we have used a single $m$ value to determine the \tln. Note that one can compute another dimensionless number using $\lambda$ that is independent of the radial distance, which is simply given by $G_{4}\lambda/(G_{4}M)^{5}$. We call this as well the dimensionless \tln, which is denoted by $\Lambda$. Whether the dimensionless \tln\ under consideration is $k_{2}$ or $\Lambda$ should be clear from the context. 

To provide another independent derivation of the same we provide a computation of the dimensionless \tln\ $k_{2}$ as well in the asymptotic limit, which reads,
%%%%%%%%%%%%%%%%%%%%%%%%%%%%%%%%%%%%%%%%%%%%%%%%%%%%%%%%%%%%%%
\begin{align}\label{TLN_v2_47}
k_{2}=\frac{1}{2}\left(\frac{2H-rH'}{3H+rH'}\right)
\end{align}
%%%%%%%%%%%%%%%%%%%%%%%%%%%%%%%%%%%%%%%%%%%%%%%%%%%%%%%%%%%%%%
The value for $H(r)$ in the asymptotic limit has already been computed, whose derivative is also simple enough to determine. Hence one can immediately find out the combinations $2H-rH'$ and $3H+rH'$ respectively. Substituting both of these expression into \ref{TLN_v2_47}, leads to the following expression for $k_{2}$,
%%%%%%%%%%%%%%%%%%%%%%%%%%%%%%%%%%%%%%%%%%%%%%%%%%%%%%%%%%%%%%
\begin{align}\label{TLN_v2_50}
k_{2}=\frac{1}{2}\left(1+\beta\right)^{5/2}
\left\lbrace -\frac{\Gamma(3-\bar{\beta})}{45\sqrt{\pi}\Gamma(-2-\bar{\beta})} 
+\frac{B}{45A}\Gamma(3+\bar{\beta})\Gamma(3-\bar{\beta}) \right\rbrace
\left(\frac{G_{4}M}{r}\right)^{5}
\end{align}
%%%%%%%%%%%%%%%%%%%%%%%%%%%%%%%%%%%%%%%%%%%%%%%%%%%%%%%%%%%%%%
This exactly coincides with \ref{TLN_v2_N07}, with the identification, $k_{2}=(3/2)G_{4}\lambda r^{-5}$. Thus the above result provides yet another verification of the derivation of \tln\ for braneworld black holes. 

%%%%%%%%%%%%%%%%%%%%%%%%%%%%%
%%%%%%%%%%%%%%%%%%%%%%%%%%%%%
%%%%%%%%%%%%%%%%%%%%%%%%%%%%%
%%%%%%%%%%%%%%%%%%%%%%%%%%%%%
\begin{figure}[ht]
\includegraphics[scale=0.8]{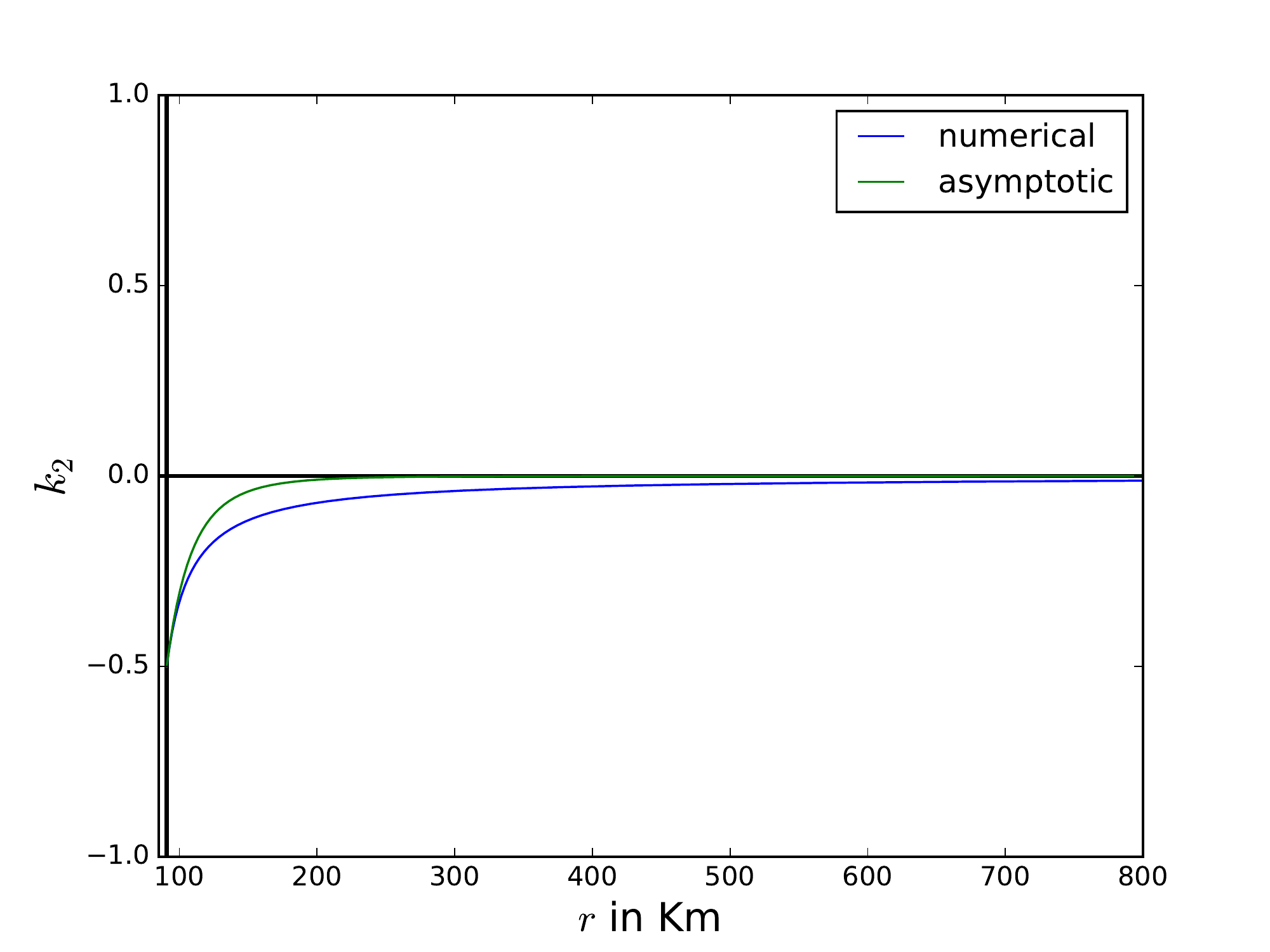}
\caption{The above figure depicts a comparison between variation of the dimensionless \tln\ $k_{2}$ with the radial distance from the black hole, derived from numerical as well as theoretical analysis. The above plots are for a black hole with mass $M_{\rm BH}=10M_{\odot})$, where $M_{\odot}$ denotes the solar mass. As evident from both theoretical and numerical analysis, the dimensionless \tln\ $k_{2}$ is non-zero and negative for any finite $r$, unlike the case for a \gr\ black hole, where it always vanishes. Further the difference between theoretical and numerical analysis is very less in the large radius limit, however in the near horizon region there is a moderate difference between them.}
\label{k2radius}
\end{figure}
%%%%%%%%%%%%%%%%%%%%%%%%%%%%%
%%%%%%%%%%%%%%%%%%%%%%%%%%%%%
%%%%%%%%%%%%%%%%%%%%%%%%%%%%%
%%%%%%%%%%%%%%%%%%%%%%%%%%%%%

%%%%%%%%%%%%%%%%%%%%%%%%%%%%%
%%%%%%%%%%%%%%%%%%%%%%%%%%%%%
%%%%%%%%%%%%%%%%%%%%%%%%%%%%%
%%%%%%%%%%%%%%%%%%%%%%%%%%%%%
\begin{figure}[t!]
\includegraphics[scale=0.8]{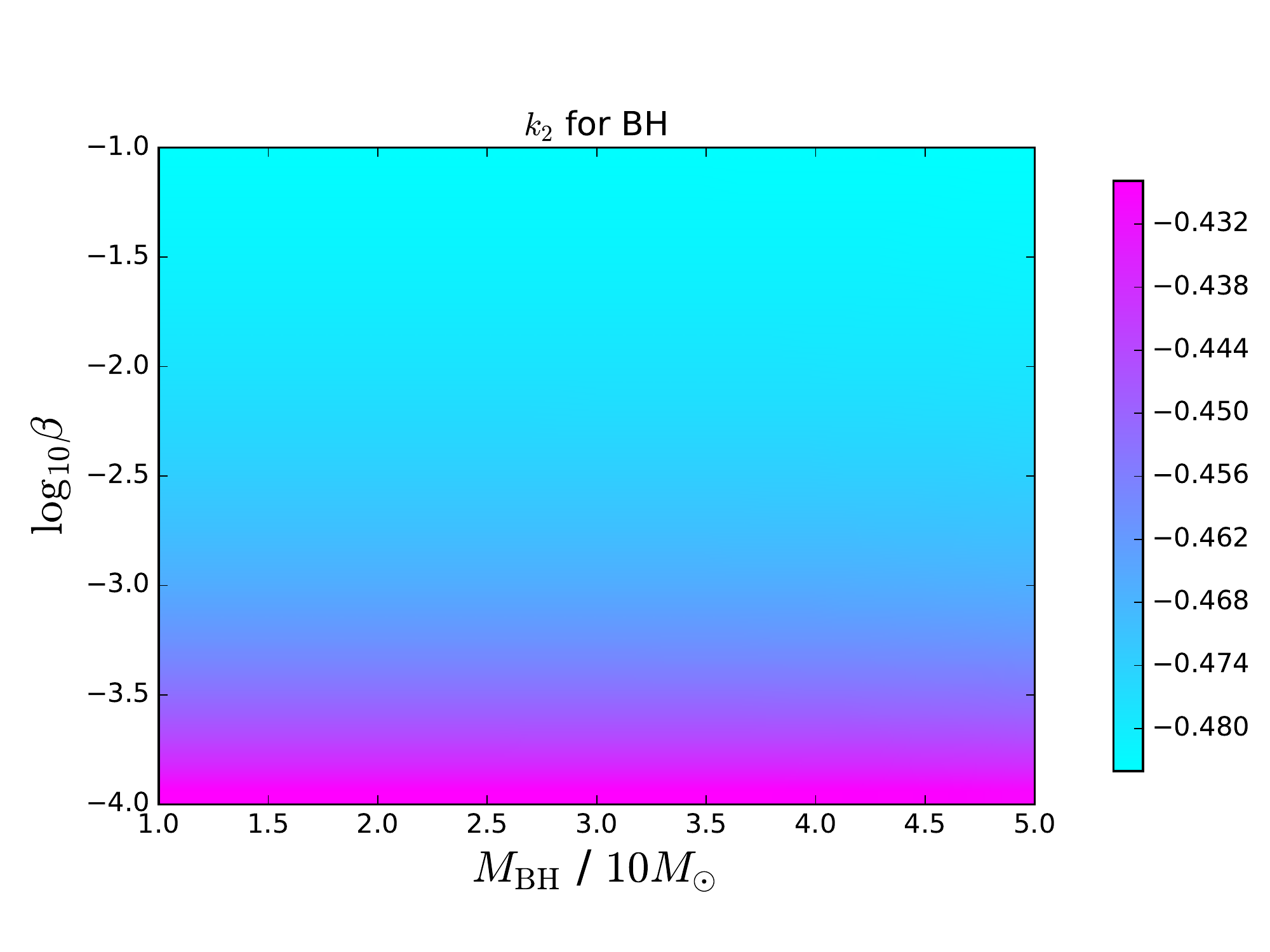}
\caption{The above figure depicts variation of the dimensionless \tln\ $k_{2}$ with the black hole mass as well as the parameter $\beta$ inherited from the extra dimensions from numerical analysis. The black hole mass $M_{\rm BH}$ has been normalized to $(M_{\rm BH}/10M_{\odot})$, where $M_{\odot}$ denotes the solar mass. As evident from the panel on the right, the dimensionless \tln\ $k_{2}$ is non-zero and negative, which is expected from our theoretical analysis as well. As evident there is very little variation of $k_{2}$ with mass but it varies considerably with $\beta$. As beta changes from $10^{-4}$ to $0.1$ the dimensionless \tln\ changes from $-0.430$ to $-0.480$. See text for more discussion.}
\label{tlnBH}
\end{figure}
%%%%%%%%%%%%%%%%%%%%%%%%%%%%%
%%%%%%%%%%%%%%%%%%%%%%%%%%%%%
%%%%%%%%%%%%%%%%%%%%%%%%%%%%%
%%%%%%%%%%%%%%%%%%%%%%%%%%%%%

%%%%%%%%%%%%%%%%%%%%%%%%%%%%%
%%%%%%%%%%%%%%%%%%%%%%%%%%%%%
%%%%%%%%%%%%%%%%%%%%%%%%%%%%%
%%%%%%%%%%%%%%%%%%%%%%%%%%%%%
\begin{figure}[t!]
\includegraphics[scale=0.8]{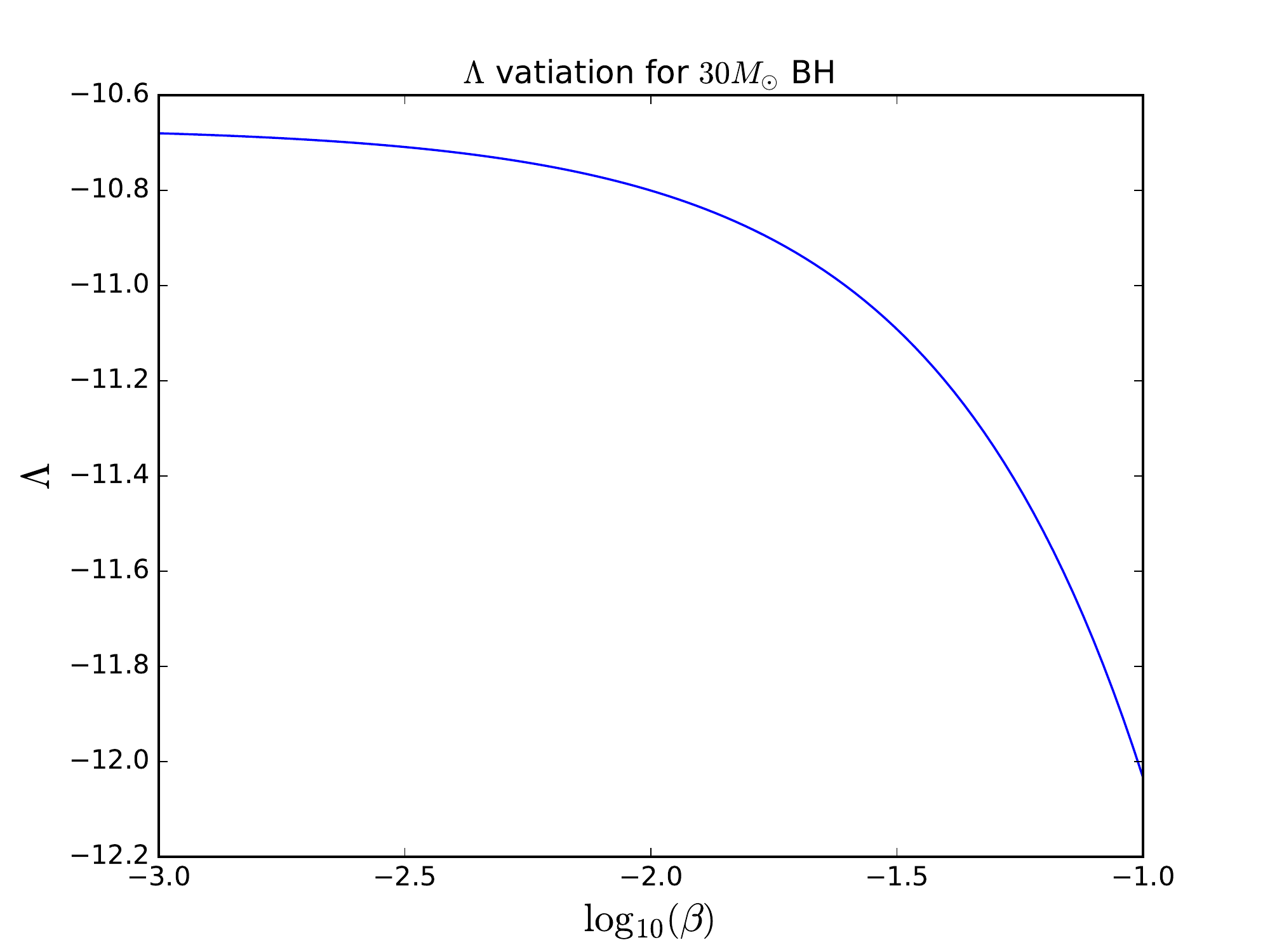}
\caption{The above figure depicts variation of the dimensionless \tln\ $\Lambda$ with the parameter $\beta$ inherited from the extra dimensions using numerical analysis. The black hole mass $M_{\rm BH}$ has been taken to be $\sim 30M_{\odot}$, where $M_{\odot}$ denotes the solar mass. As evident, alike $k_{2}$, the other dimensionless \tln\ $\Lambda$ is also non-zero and negative, which is expected from our theoretical analysis as well. There is little variation of $\Lambda$ with $\beta$, which is $\sim 20\%$. As beta changes from $10^{-3}$ to $0.1$, the dimensionless \tln\ $\Lambda$ changes from $-10.6$ to $-12.2$.}
\label{Lambdabh}
\end{figure}
%%%%%%%%%%%%%%%%%%%%%%%%%%%%%
%%%%%%%%%%%%%%%%%%%%%%%%%%%%%
%%%%%%%%%%%%%%%%%%%%%%%%%%%%%
%%%%%%%%%%%%%%%%%%%%%%%%%%%%%

However in order to find out the numerical estimation of the dimensionless \tln, we need to determine the unknown constants $A$ and $B$ appearing in \ref{TLN_v2_N07} as well as in \ref{TLN_v2_50}. These unknown constants can be obtained by using the gravitational perturbation $H(r)$ and its derivative $H'(r)$ at some fixed radius $R$. For black holes this would correspond to the event horizon, while for neutron star this is the radius determining the extent of the neutron star. Since for \gr, an exact solution of the perturbation equation is known, it is possible to explicitly determine the \tln\ in terms of these quantities. However, unlike the situation in \gr, in the present context we do not have a handle on the analytic solution for the gravitational perturbation $H(r)$ at an arbitrary radius. Thus we cannot compute $A$ and $B$ using $H(r)$ and $rH'(r)$ at some finite radius $R$. But some physical insight may be gained by looking at the expression for the dimensionless \tln\ $\Lambda$ instead, for which the following expression can be determined from \ref{TLN_v2_N07} as, 
%%%%%%%%%%%%%%%%%%%%%%%%%%%%%%%%%%%%%%%%%%%%%%%%%%%%%%%%%%%%%%
\begin{align}\label{TLN_v2_51}
\Lambda=\frac{B}{135 A}\Bigg\lbrace \left(1+\beta\right)^{5/2}\Gamma(3+\bar{\beta})
\Gamma(3-\bar{\beta})\Bigg\rbrace
-\Bigg\lbrace \frac{\Gamma(3-\bar{\beta})}{\Gamma(-2-\bar{\beta})}\frac{\left(1+\beta\right)^{5/2}}{135\sqrt{\pi}}\Bigg\rbrace \end{align}
%%%%%%%%%%%%%%%%%%%%%%%%%%%%%%%%%%%%%%%%%%%%%%%%%%%%%%%%%%%%%%
As evident, vanishing of $\beta$ implies $\bar{\beta}\rightarrow 2$ and hence a $\Gamma(-4)$ term, which diverges, comes into existence in the denominator of the second quantity in the above expression. Thus we indeed have $\Lambda \rightarrow \Lambda _{\rm GR}=(8B/45A)$ as $\beta$ vanishes. Hence the desired general relativistic limit can always be arrived at. However, like for $k_{2}$, for $\Lambda$ as well it is difficult to comment on its nature analytically due to presence of the arbitrary constants $A$ and $B$. As evident from \ref{TLN_v2_51}, in the presence of higher dimensions, it is most likely that even for black holes the dimensionless \tln\ $\Lambda$ will be non-zero. Furthermore, depending on the sign of the ratio $(B/A)$, the dimensionless \tln\ $\Lambda$ can also turn negative. If correct, such a non-zero \emph{but} negative value of the dimensionless tidal Love number $\Lambda$ may serve as a distinct signature of the existence of higher dimensions.

At this stage it will be worthwhile to point out implications of the above result for $k_{2}$, the other dimensionless \tln\ presented in \ref{TLN_v2_50}. Since the dimensionless \tln\ $k_{2}$ varies with radius as $r^{-5}$, it follows that asymptotically it should vanish. In the context of \gr, on the other hand, $k_{2}$ identically vanishes everywhere, not just asymptotically. While in presence of extra dimensions, following the above discussion, one could possibly argue that the quantity $(B/A)$ may result into a negative contribution to the dimensionless \tln\ $k_{2}$, which ultimately asymptote to zero. 

The analytical understanding regarding the dimensionless \tln\ $k_{2}$ or $\Lambda$ presented above, can also be verified by numerically solving the general differential equation, presented in \ref{TLN_v2_39}. The result of such a numerical analysis has been presented in \ref{k2radius}, where variation of the dimensionless \tln\ $k_{2}$ with radial distance from black hole has been depicted. Similarly in \ref{tlnBH} numerical estimates of the dimensionless \tln\ $k_{2}$ have been plotted against the parameter $\beta$ and the black hole mass. The plots presented in both \ref{k2radius} and in \ref{tlnBH} clearly reveals a negative value for the dimensionless \tln\ $k_{2}$, further bolstering our claim. This has been achieved by the computation of $y(\partial _{y}H/H)$ starting from the differential equation presented in \ref{TLN_v2_39}, as $y$ approaches the asymptotic limit. The numerical computation reveals that for all values of $y$, even when it reaches some asymptotic limit, $y(\partial _{y}H/H)$ remains above $2$ and hence from \ref{TLN_v2_47} it immediately follows that $k_{2}$ should be negative (see \ref{k2radius} for a clear illustration of this fact). Further, the correctness of the theoretical techniques demonstrated above is also evident by the comparison of theoretical estimations with numerical analysis as presented in \ref{k2radius}. Since the match is almost exact at large radial distance we have computed $k_{2}$ at $r=400~\textrm{km}$ in the plot presented in \ref{tlnBH}. Thus the negativity and non-zero nature of the dimensionless \tln\ in this situation is borne out by our numerical simulations as well (see \ref{Lambdabh} for variation of the other dimensionless tidal Love number $\Lambda$ with $\beta$). 

Before concluding this section, let us briefly mention that even though these results are derived in the context of black holes, they remain equally applicable in the context of neutron stars as well, as long as the Weyl stress in the exterior region of the neutron star continues to be given by \ref{TLN_v1_27}. In that case, \ref{TLN_v2_50} still provides the tidal Love number, but the unknown constants need to be determined on the surface of the star. But that requires an understanding of the gravitational perturbation in the interior of the neutron star. In the next section we explicitly demonstrate how to achieve the same. 
%%%%%%%%%%%%%%%%%%%%%%%%%%%%%%%%%%%%%%%%%%%%%%%%%%%%%%%%%%%%%%%%%%%%%%%%%%%%%%%%%%%%%%%%%%%%%%%%%%%
%%%%%%%%%%%%%%%%%%%%%%%%%%%%%%%%%%%%%%%%%%%%%%%%%%%%%%%%%%%%%%%%%%%%%%%%%%%%%%%%%%%%%%%%%%%%%%%%%%%
%%%%%%%%%%%%%%%%%%%%%%%%%%%%%%%%%%%%%%%%%%%%%%%%%%%%%%%%%%%%%%%%%%%%%%%%%%%%%%%%%%%%%%%%%%%%%%%%%%%
\section{Tidal Love numbers for a neutron star on the brane}\label{TLN_v1_Neutron}

In this section we will compute the tidal Love number associated with a neutron star located on the brane. The results derived in \ref{TLN_v1_Perturbation} and \ref{TLN_v1_Black} will still remain useful albeit in the exterior of the neutron star. In particular, if we assume that the dark radiation and the dark pressure have the same form outside of the neutron star, then asymptotic behaviour of the metric perturbation remains identical. Hence \ref{TLN_v2_50} will hold in the exterior of the neutron star as well. However the determination of the unknown constants can be achieved only if the gravitational perturbation in the interior of the neutron star is known. This must be done in a separate manner as the matter equation of state parameter also comes into play. 
%%%%%%%%%%%%%%%%%%%%%%%%%%%%%%%%%%%%%%%%%%%%%%%%%%%%%%%%%%%%%%%%%%%%%
%%%%%%%%%%%%%%%%%%%%%%%%%%%%%%%%%%%%%%%%%%%%%%%%%%%%%%%%%%%%%%%%%%%%%
%%%%%%%%%%%%%%%%%%%%%%%%%%%%%%%%%%%%%%%%%%%%%%%%%%%%%%%%%%%%%%%%%%%%%
\subsection{Background equations in the interior of the neutron star}\label{TLN_v1_NeutronBack}

Let us discuss the background spacetime in the interior of the neutron star in presence of an extra dimension. The presence of matter in the interior of the neutron star complicates the situation significantly by introducing linear as well as quadratic terms depending on the matter energy momentum tensor. In particular, the effective gravitational field equations in this context take the following form \cite{Maartens:2003tw},
%%%%%%%%%%%%%%%%%%%%%%%%%%%%%%%%%%%%%%%%%%%%%%%%%%%%%%%%%%%%%%
\begin{align}\label{TLN_v3_02}
G_{\mu \nu}+E_{\mu \nu}=8\pi G_{4}T_{\mu \nu}+\left(8\pi G_{5}\right)^{2}\left(-\frac{1}{4}T_{\mu \alpha}T^{\alpha}_{\nu}+\frac{1}{12}TT_{\mu \nu}+\frac{1}{8}g_{\mu \nu}T_{\alpha \beta}T^{\alpha \beta}-\frac{1}{24}T^{2}g_{\mu \nu}\right)~,
\end{align}
%%%%%%%%%%%%%%%%%%%%%%%%%%%%%%%%%%%%%%%%%%%%%%%%%%%%%%%%%%%%%%
where $G_{5}$ is the five dimensional gravitational constant, $T_{\mu \nu}$ is the matter energy momentum tensor and $E_{\mu \nu}$ is the projection of the Weyl tensor introduced above. To proceed further, we will assume that $E_{\mu \nu}$ can be represented as in \ref{TLN_v1_Perturbation}, while $T_{\mu \nu}$ depicts a perfect fluid, with some energy density $\rho$ and pressure $p$. Thus for a static and spherically symmetric background spacetime, whose line element is given by \ref{TLN_v1_02}, the associated field equations take the following form,
%%%%%%%%%%%%%%%%%%%%%%%%%%%%%%%%%%%%%%%%%%%%%%%%%%%%%%%%%%%%%%
\begin{align}
e^{-2\lambda(r)}\left(\frac{1}{r^{2}}-\frac{2\lambda'}{r}\right)-\frac{1}{r^{2}}&=-8\pi G_{4}\rho\left(1+\frac{\rho}{2\lambda _{\rm b}}\right)
-24\pi \tilde{U}(r)
\label{TLN_v3_03a}
\\
e^{-2\lambda(r)}\left(\frac{2\nu'}{r}+\frac{1}{r^{2}}\right)-\frac{1}{r^{2}}&=8\pi G_{4}\left\{p+\frac{\rho}{2\lambda _{\rm b}}\left(\rho+2p\right) \right\}+8\pi \left(\tilde{U}+2\tilde{P}\right)
\label{TLN_v3_03b}
\\
e^{-2\lambda}\left\{\nu''+\nu'^{2}-\nu'\lambda'+\frac{1}{r}\left(\nu'-\lambda'\right)\right\}&=16\pi G_{4}\left\{p+\frac{\rho}{2\lambda _{\rm b}}\left(\rho+2p\right) \right\}+16\pi\left(\tilde{U}-\tilde{P}\right)
\label{TLN_v3_03c}
\end{align}
%%%%%%%%%%%%%%%%%%%%%%%%%%%%%%%%%%%%%%%%%%%%%%%%%%%%%%%%%%%%%%
Here we have defined $\tilde{U}$ and $\tilde{P}$ as in \ref{TLN_v1_04} with $G_{4}$ being the four dimensional Newton's constant and $\lambda _{\rm b}$ is the brane tension. Thus the above set of equations can be thought of as the background equations associated with an anisotropic fluid with two additional components: one coming from the bulk Weyl tensor (i.e., dark radiation and dark pressure components) characterized by $\tilde{U}$ and $\tilde{P}$, and one coming from matter energy momentum tensor and its quadratic combination. In this case, the effective energy density $\rho_{\rm eff}$ and effective pressure $p_{\rm eff}$ are 
%%%%%%%%%%%%%%%%%%%%%%%%%%%%%%%%%%%%%%%%%%%%%%%%%%%%%%%%%%%%%%
\begin{align}\label{TLN_v3_05}
\rho _{\rm eff}=\rho\left(1+\frac{\rho}{2\lambda _{b}}\right);
\qquad 
p_{\rm eff}=p+\frac{\rho}{2\lambda _{\rm b}}\left(\rho+2p\right)
\end{align}
%%%%%%%%%%%%%%%%%%%%%%%%%%%%%%%%%%%%%%%%%%%%%%%%%%%%%%%%%%%%%%
Hence the above set of equations depict an anisotropic two-fluid system as the background. Thus in the computation of gravitational perturbation around this background, we not only have to consider perturbations of $\tilde{U}$ and $\tilde{P}$, but also of $\rho$ and $p$. As we will see, this will immediately lead to problems by complicating the scenario quite a bit. As it will turn out, even then we can make some educated choice about the nature of Weyl fluid in the interior of the neutron star to go around the issues of the two fluid system. 
%%%%%%%%%%%%%%%%%%%%%%%%%%%%%%%%%%%%%%%%%%%%%%%%%%%%%%%%%%%%%%%%%%%%%
%%%%%%%%%%%%%%%%%%%%%%%%%%%%%%%%%%%%%%%%%%%%%%%%%%%%%%%%%%%%%%%%%%%%%
%%%%%%%%%%%%%%%%%%%%%%%%%%%%%%%%%%%%%%%%%%%%%%%%%%%%%%%%%%%%%%%%%%%%%
\subsection{Perturbation equations in the interior of the neutron star}\label{TLN_v1_NeutronInt}

The gravitational perturbation equations in the exterior of the neutron star has already been discussed in the earlier sections. In this section we will concentrate on the gravitational perturbation in the interior of the neutron star. The modifications to the gravitational field equations due to presence of matter has already been demonstrated in \ref{TLN_v3_02}. The gravitational perturbation inside the neutron star in a static and spherically symmetric background obeys \ref{TLN_v3_03a}--\ref{TLN_v3_03c} and involves both the bulk Weyl tensor and linear as well as quadratic contributions from the matter sector. 

In this context as well, the fact that perturbations must satisfy the symmetries of the background spacetime, brings the even parity perturbation to the form advocated in \ref{TLN_v3_01}. Since the angular component of the energy momentum tensor of the matter fluid as well as the contribution from bulk Weyl tensor are identical, it follows that $\delta G^{\theta}_{\theta}-\delta G^{\phi}_{\phi}=0$. Hence, as in \ref{TLN_v1_Even}, in this case also we will have $H_{0}=H_{2}\equiv H(r)$. Similarly, the result $\delta G^{r}_{\theta}=8\pi G_{4}\delta T^{r}_{\theta}$ will lead to $K'=H'+2\nu'H$, which is identical to the one obtained in \ref{TLN_v1_Even}. Thus the fact that only two perturbation components ($H(r)$ and $K(r)$) are necessary to characterize the even parity gravitational perturbation and that they are connected by a differential equation holds both inside and outside the neutron star. On the other hand, all the remaining relations connecting components of the perturbed metric with perturbations in the matter sector will lead to a different set of equations. Let us start with the addition of the perturbation equations in the angular directions, i.e., let us concentrate on the equation $\delta G^{\theta}_{\theta}+\delta G^{\phi}_{\phi}=8\pi G_{4}(\delta T^{\theta}_{\theta}+\delta T^{\phi}_{\phi})$. On expanding the above equation, we will have the following structure,
%%%%%%%%%%%%%%%%%%%%%%%%%%%%%%%%%%%%%%%%%%%%%%%%%%%%%%%%%%%%%%
\begin{align}\label{TLN_v1_38}
e^{-2\lambda}r^{2}K''&+e^{-2\lambda}rK'\left\{2+r\left(\nu'-\lambda'\right)\right\}-e^{-2\lambda}r^{2}H''
-e^{-2\lambda}rH'\left(3r\nu'-r\lambda'+2\right)
\nonumber
\\
&-16\pi r^{2}\left(\frac{\partial \tilde{U}}{\partial \tilde{P}}-1\right)\delta \tilde{P}-16\pi r^{2}H\left(\tilde{U}-\tilde{P}\right)
-16\pi r^{2}H\tilde{p}_{\rm eff}-16\pi r^{2}\delta \tilde{p}_{\rm eff}=0
\end{align}
%%%%%%%%%%%%%%%%%%%%%%%%%%%%%%%%%%%%%%%%%%%%%%%%%%%%%%%%%%%%%%
Here $\tilde{U}$ and $\tilde{P}$ has the usual expressions and $\tilde{\rho}_{\rm eff}=G_{4}\rho_{\rm eff}$ and $\tilde{p}_{\rm eff}=G_{4}p_{\rm eff}$. Subsequently, the radial perturbation equation, namely $\delta G^{r}_{r}=8\pi G_{4} \delta T^{r}_{r}$ yields,
%%%%%%%%%%%%%%%%%%%%%%%%%%%%%%%%%%%%%%%%%%%%%%%%%%%%%%%%%%%%%%
\begin{align}\label{TLN_v1_39}
e^{-2\lambda}\left(1+r\nu'\right)rK'&-\left\{\frac{1}{2}\ell\left(\ell+1\right)-1\right\}K-e^{-2\lambda}rH'
+\left\{\frac{1}{2}\ell\left(\ell+1\right)-1-8\pi r^{2}\left(\tilde{p}_{\rm eff}+\tilde{U}+2\tilde{P}\right)\right\}H
\nonumber
\\
&-8\pi r^{2}\delta \tilde{p}_{\rm eff}-8\pi r^{2}\left(2+\frac{\partial \tilde{U}}{\partial \tilde{P}}\right)\delta \tilde{P}=0
\end{align}
%%%%%%%%%%%%%%%%%%%%%%%%%%%%%%%%%%%%%%%%%%%%%%%%%%%%%%%%%%%%%%
Finally, the remaining equation corresponding to the temporal part of the perturbation equation, namely $\delta G^{t}_{t}=8\pi G_{4} \delta T^{t}_{t}$ results into,
%%%%%%%%%%%%%%%%%%%%%%%%%%%%%%%%%%%%%%%%%%%%%%%%%%%%%%%%%%%%%%
\begin{align}\label{TLN_v1_40}
e^{-2\lambda}r^{2}K''&+e^{-2\lambda}rK'\left(3-r\lambda'\right)-\left\{\frac{1}{2}\ell (\ell+1)-1\right\}K
-re^{-2\lambda}H'-\left\{\frac{1}{2}\ell\left(\ell+1\right)+1-8\pi r^{2}\left(\tilde{\rho}_{\rm eff}+3\tilde{U}\right)\right\}H
\nonumber
\\
&+8\pi r^{2}\left(\frac{\partial \tilde{\rho}_{\rm eff}}{\partial \tilde{p}_{\rm eff}}\right)\delta \tilde{p}_{\rm eff}
+8\pi r^{2}\left(\frac{\partial \tilde{U}}{\partial \tilde{P}} \right)\delta \tilde{P} =0
\end{align}
%%%%%%%%%%%%%%%%%%%%%%%%%%%%%%%%%%%%%%%%%%%%%%%%%%%%%%%%%%%%%%
The difficulty associated with the above equations, should be apparent by now. Unlike the perturbation equations in the exterior of the neutron star, here we have to eliminate both $\delta \tilde{P}$ and $\delta \tilde{p}_{\rm eff}$. This, as we will see, will result into terms depending on $K$. Hence it will not be possible, in the most general setting to arrive at a single master equation, governing all the even parity gravitational perturbations (for a similar situation, see \cite{Char:2018grw}). 

However, for the moment being let us press further and see how far we can reach. For that purpose, we start by solving for $\delta \tilde{p}_{\rm eff}$, which can be done using \ref{TLN_v1_38} leading to, 
 %%%%%%%%%%%%%%%%%%%%%%%%%%%%%%%%%%%%%%%%%%%%%%%%%%%%%%%%%%%%%%
\begin{align}\label{TLN_v1_41}
16\pi r^{2}\delta \tilde{p}_{\rm eff}&=e^{-2\lambda}r^{2}K''+e^{-2\lambda}rK'\left\{2+r\left(\nu'-\lambda'\right)\right\}-e^{-2\lambda}r^{2}H''
-e^{-2\lambda}rH'\left(3r\nu'-r\lambda'+2\right)
\nonumber
\\
&-16\pi r^{2}\left(\frac{\partial \tilde{U}}{\partial \tilde{P}}-1\right)\delta \tilde{P}-16\pi r^{2}H\left(\tilde{U}-\tilde{P}\right)
-16\pi r^{2}H\tilde{p}_{\rm eff}
\end{align}
%%%%%%%%%%%%%%%%%%%%%%%%%%%%%%%%%%%%%%%%%%%%%%%%%%%%%%%%%%%%%%
Thus using \ref{TLN_v1_41} we can replace the $\delta \tilde{p}_{\rm eff}$ in the other two equations. However, they both will depend on $\delta \tilde{P}$. Hence, to eliminate $\delta \tilde{P}$ we focus on \ref{TLN_v1_39}. First of all we substitute $\delta \tilde{p}_{\rm eff}$ from the above equation into \ref{TLN_v1_39}, which results into the following expression,
%%%%%%%%%%%%%%%%%%%%%%%%%%%%%%%%%%%%%%%%%%%%%%%%%%%%%%%%%%%%%%
\begin{align}\label{TLN_v1_42}
e^{-2\lambda}\left(1+r\nu'\right)rK'&-\left\{\frac{1}{2}\ell\left(\ell+1\right)-1\right\}K-e^{-2\lambda}rH'
+\left\{\frac{1}{2}\ell\left(\ell+1\right)-1-8\pi r^{2}\left(\tilde{p}_{\rm eff}+\tilde{U}+2\tilde{P}\right)\right\}H
\nonumber
\\
&-8\pi r^{2}\left(2+\frac{\partial \tilde{U}}{\partial \tilde{P}}\right)\delta \tilde{P}
=\frac{1}{2}e^{-2\lambda}r^{2}K''+\frac{1}{2}e^{-2\lambda}rK'\left\{2+r\left(\nu'-\lambda'\right)\right\}
\nonumber
\\
&-\frac{1}{2}e^{-2\lambda}r^{2}H''
-\frac{1}{2}e^{-2\lambda}rH'\left(3r\nu'-r\lambda'+2\right)
\nonumber
\\
&-8\pi r^{2}\left(\frac{\partial \tilde{U}}{\partial \tilde{P}}-1\right)\delta \tilde{P}-8\pi r^{2}H\left(\tilde{U}-\tilde{P}\right)
-8\pi r^{2}H\tilde{p}_{\rm eff}
\end{align}
%%%%%%%%%%%%%%%%%%%%%%%%%%%%%%%%%%%%%%%%%%%%%%%%%%%%%%%%%%%%%%
After simplifying the above equation further one can immediately solve for the quantity $\delta \tilde{P}$, leading to,
%%%%%%%%%%%%%%%%%%%%%%%%%%%%%%%%%%%%%%%%%%%%%%%%%%%%%%%%%%%%%%
\begin{align}\label{TLN_v1_44}
48\pi r^{2}\delta \tilde{P}&=-48\pi r^{2}H\tilde{P}-2\left\{\frac{1}{2}\ell\left(\ell+1\right)-1\right\}K
+2\left\{\frac{1}{2}\ell\left(\ell+1\right)-1\right\}H
\nonumber
\\
&-e^{-2\lambda}r^{2}K''+e^{-2\lambda}r^{2}K'\left(\nu'+\lambda'\right)
+e^{-2\lambda}r^{2}H''-e^{-2\lambda}rH'\left(3r\nu'-r\lambda'\right)
\end{align}
%%%%%%%%%%%%%%%%%%%%%%%%%%%%%%%%%%%%%%%%%%%%%%%%%%%%%%%%%%%%%%
In the final step one has to use both \ref{TLN_v1_41} and \ref{TLN_v1_44} to eliminate both $\delta \tilde{p}$ and $\delta \tilde{P}$ in \ref{TLN_v1_40}, which can be done in a straightforward manner. However, the differential equation so obtained will have dependance on $K(r)$ along with its derivative. Thus unlike the exterior scenario, even if one use the relation $K'=H'+2\nu'H$, all the dependences on the angular part of gravitational perturbation can not be eliminated. This suggests that in the interior one needs to solve a set of coupled differential equations. This makes handling the interior structure of the neutron star in presence of extra dimensions more difficult to work with. 

However, one can work around this problem and arrive at interesting scenarios having an exact solution to the above problem if some suitable assumptions are being made. Before going into the details, note that the problem of getting a closed form solution is mainly associated with the fact that in the interior of the neutron star we have a two fluid system. If we can set the extra dimensional contribution coming through the dark radiation and dark pressure to zero, we may have a possibility to circumvent the problem. In particular, the quantity of significant interest in this context corresponds to the continuity of matter and metric across the surface of the neutron star. This continuity equation in presence of extra dimension reads \cite{Germani:2001du},
%%%%%%%%%%%%%%%%%%%%%%%%%%%%%%%%%%%%%%%%%%%%%%%%%%%%%%%%%%%%%%
\begin{align}\label{TLN_v3_07}
\tilde{p}_{\rm eff}+\tilde{U}_{-}+2\tilde{P}_{-}=\tilde{U}_{+}+2\tilde{P}_{+}=\frac{3\tilde{P}_{0}}{2R^{4}}
\end{align}
%%%%%%%%%%%%%%%%%%%%%%%%%%%%%%%%%%%%%%%%%%%%%%%%%%%%%%%%%%%%%%
where $R$ is the radius of the neutron star. Here, the `+' sign corresponds to configuration in the exterior of the neutron star, while `-' sign is associated with the interior configuration. Thus the most economic way to get an analytic handle on the perturbation equation assumes that $\tilde{U}_{-}=0=\tilde{P}_{-}$, i.e., the effects from extra dimension due to the Weyl stress identically vanishes at the interior of the neutron star. This simplifies the analysis presented above significantly and more importantly reduces the above system of equations representing gravitational perturbation in the interior of the neutron star to a single fluid system. In particular, the field equations presented in \ref{TLN_v1_38}--\ref{TLN_v1_40} can be rewritten in the following manner, 
%%%%%%%%%%%%%%%%%%%%%%%%%%%%%%%%%%%%%%%%%%%%%%%%%%%%%%%%%%%%%%
\begin{align}
e^{-2\lambda}r^{2}K''&+e^{-2\lambda}rK'\left\{2+r\left(\nu'-\lambda'\right)\right\}-e^{-2\lambda}r^{2}H''
-e^{-2\lambda}rH'\left(3r\nu'-r\lambda'+2\right)
\nonumber
\\
&-16\pi r^{2}H\tilde{p}_{\rm eff}-16\pi r^{2}\delta \tilde{p}_{\rm eff}=0
\label{TLN_v3_08a}
\\
e^{-2\lambda}\Big(1&+r\nu'\Big)rK'-\left\{\frac{1}{2}\ell\left(\ell+1\right)-1\right\}K-e^{-2\lambda}rH'
+\left\{\frac{1}{2}\ell\left(\ell+1\right)-1-8\pi r^{2}\tilde{p}_{\rm eff}\right\}H
\nonumber
\\
&-8\pi r^{2}\delta \tilde{p}_{\rm eff}=0
\label{TLN_v3_08b}
\\
e^{-2\lambda}r^{2}K''&+e^{-2\lambda}rK'\left(3-r\lambda'\right)-\left\{\frac{1}{2}\ell (\ell+1)-1\right\}K
-\left\{\frac{1}{2}\ell\left(\ell+1\right)+1-8\pi r^{2}\tilde{\rho}_{\rm eff}\right\}H
\nonumber
\\
&-re^{-2\lambda}H'+8\pi r^{2}\left(\frac{\partial \tilde{\rho}_{\rm eff}}{\partial \tilde{p}_{\rm eff}}\right)\delta \tilde{p}_{\rm eff}=0
\label{TLN_v3_08c}
\end{align}
%%%%%%%%%%%%%%%%%%%%%%%%%%%%%%%%%%%%%%%%%%%%%%%%%%%%%%%%%%%%%%
As emphasized earlier, the gravitational perturbation equations now involves only a single matter field with energy density $\tilde{\rho}_{\rm eff}$ and isotropic pressure $\tilde{p}_{\rm eff}$ along with their perturbations. Note that even though the influence of the Weyl fluid vanishes in the interior, the effect of extra dimension is still present through the effective matter energy momentum tensor. Since the above equations reduce to the familiar form as in \ref{TLN_v1_Perturbation}, we can employ the same strategy here as well, i.e., we will first eliminate any term involving $\delta \tilde{p}_{\rm eff}$ and then shall eliminate all the reference to the $K$ term appearing in these equations through the help of the relation $K'=H'+2\nu'H$. Thus finally combining all these equations in an appropriate manner we obtain a single differential equation for the master variable inside the neutron star to read,
%%%%%%%%%%%%%%%%%%%%%%%%%%%%%%%%%%%%%%%%%%%%%%%%%%%%%%%%%%%%%%
\begin{align}\label{TLN_v3_09}
H''+\Bigg[\frac{2}{r}+e^{2\lambda}\Big\{\frac{2\tilde{m}_{\rm eff}(r)}{r^{2}}&+4\pi r\left(\tilde{p}_{\rm eff}-\tilde{\rho}_{\rm eff}\right) \Big\} \Bigg]H'
\nonumber
\\
&+H\Bigg[-\ell(\ell+1)\frac{e^{2\lambda}}{r^{2}}+4\pi e^{2\lambda}\left\{5\tilde{\rho}_{\rm eff}+9\tilde{p}_{\rm eff}+\frac{\tilde{\rho}_{\rm eff}+\tilde{p}_{\rm eff}}{d\tilde{p}_{\rm eff}/d\tilde{\rho}_{\rm eff}} \right\}-4\nu'^{2}\Bigg]=0
\end{align}
%%%%%%%%%%%%%%%%%%%%%%%%%%%%%%%%%%%%%%%%%%%%%%%%%%%%%%%%%%%%%%
This is essentially a general relativity problem, but with a different energy density and pressure, which must match with the exterior solution having non-zero Weyl stress. The important point about the above differential equation is that the energy densities and pressure appearing in it are not just matter energy density and pressure, they also have contributions from the higher dimensions through the brane tension $\lambda _{\rm b}$. In particular, the equation of state parameter for the effective fluid becomes,
%%%%%%%%%%%%%%%%%%%%%%%%%%%%%%%%%%%%%%%%%%%%%%%%%%%%%%%%%%%%%%
\begin{align}\label{TLN_v3_06}
\frac{\partial \tilde{\rho}_{\rm eff}}{\partial \tilde{p}_{\rm eff}}=\frac{\partial \rho_{\rm eff}}{\partial p_{\rm eff}}=\frac{\partial \rho_{\rm eff}}{\partial p}\left(\frac{\partial p_{\rm eff}}{\partial p}\right)^{-1}
=\left(1+\frac{\rho}{\lambda _{\rm b}}\right)\frac{\partial \rho}{\partial p}\left(1+\frac{\rho}{\lambda _{\rm b}}\frac{\partial \rho}{\partial p}+\frac{\rho}{\lambda_{\rm b}}+\frac{p}{\lambda _{\rm b}}\frac{\partial \rho}{\partial p} \right)^{-1}
\end{align}
%%%%%%%%%%%%%%%%%%%%%%%%%%%%%%%%%%%%%%%%%%%%%%%%%%%%%%%%%%%%%%
Thus indeed the equation of state parameter gets modified in the presence of extra dimension through the brane tension $\lambda _{\rm b}$, such that in the limit $\lambda _{\rm b} \rightarrow 0$, we get back the correct equation of state parameter. 

%%%%%%%%%%%%%%%%%%%%%%%%%%%%%
%%%%%%%%%%%%%%%%%%%%%%%%%%%%%
%%%%%%%%%%%%%%%%%%%%%%%%%%%%%
%%%%%%%%%%%%%%%%%%%%%%%%%%%%%
\begin{figure}[t!]
\includegraphics[scale=0.41]{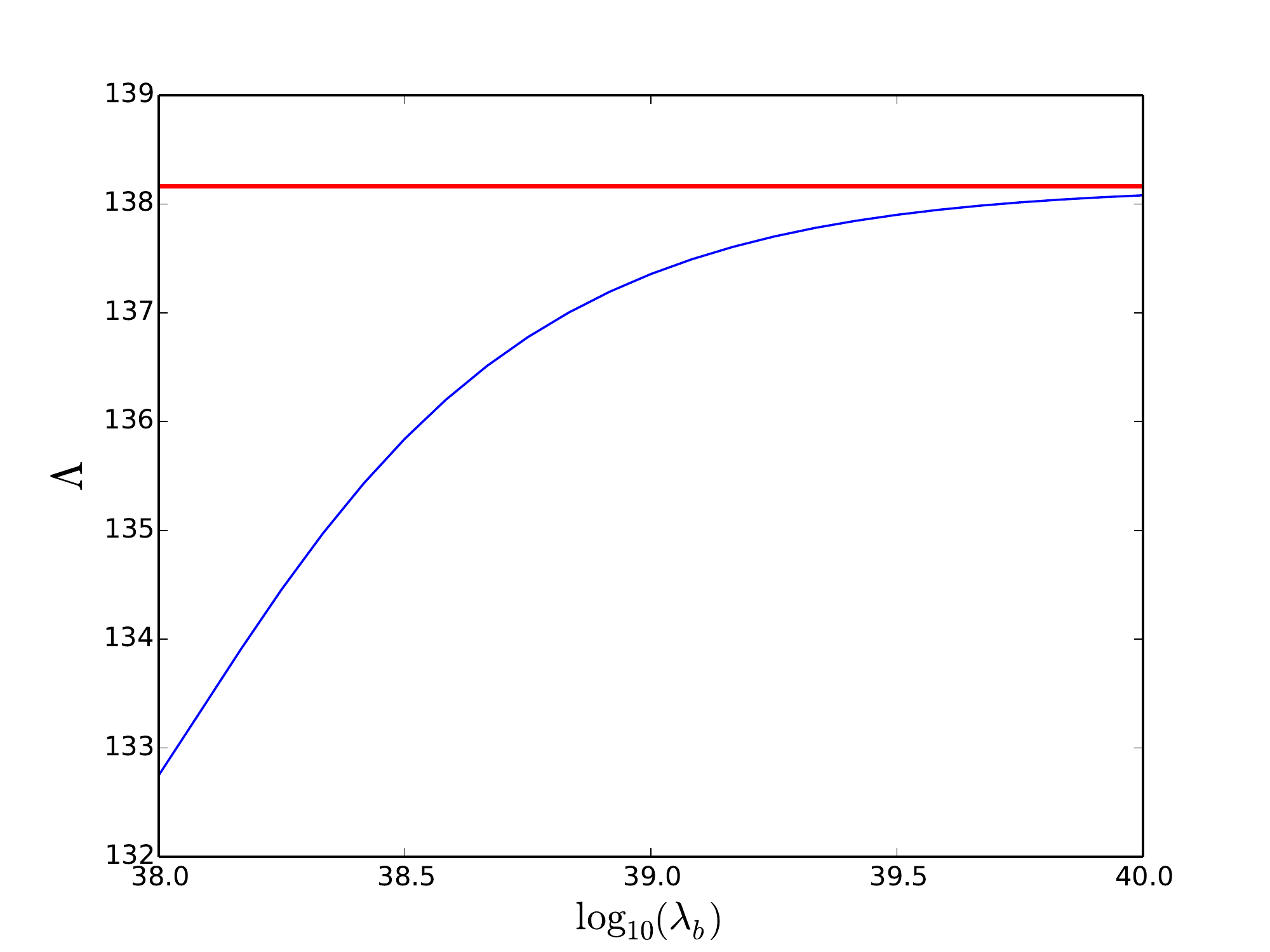}~~
\includegraphics[scale=0.41]{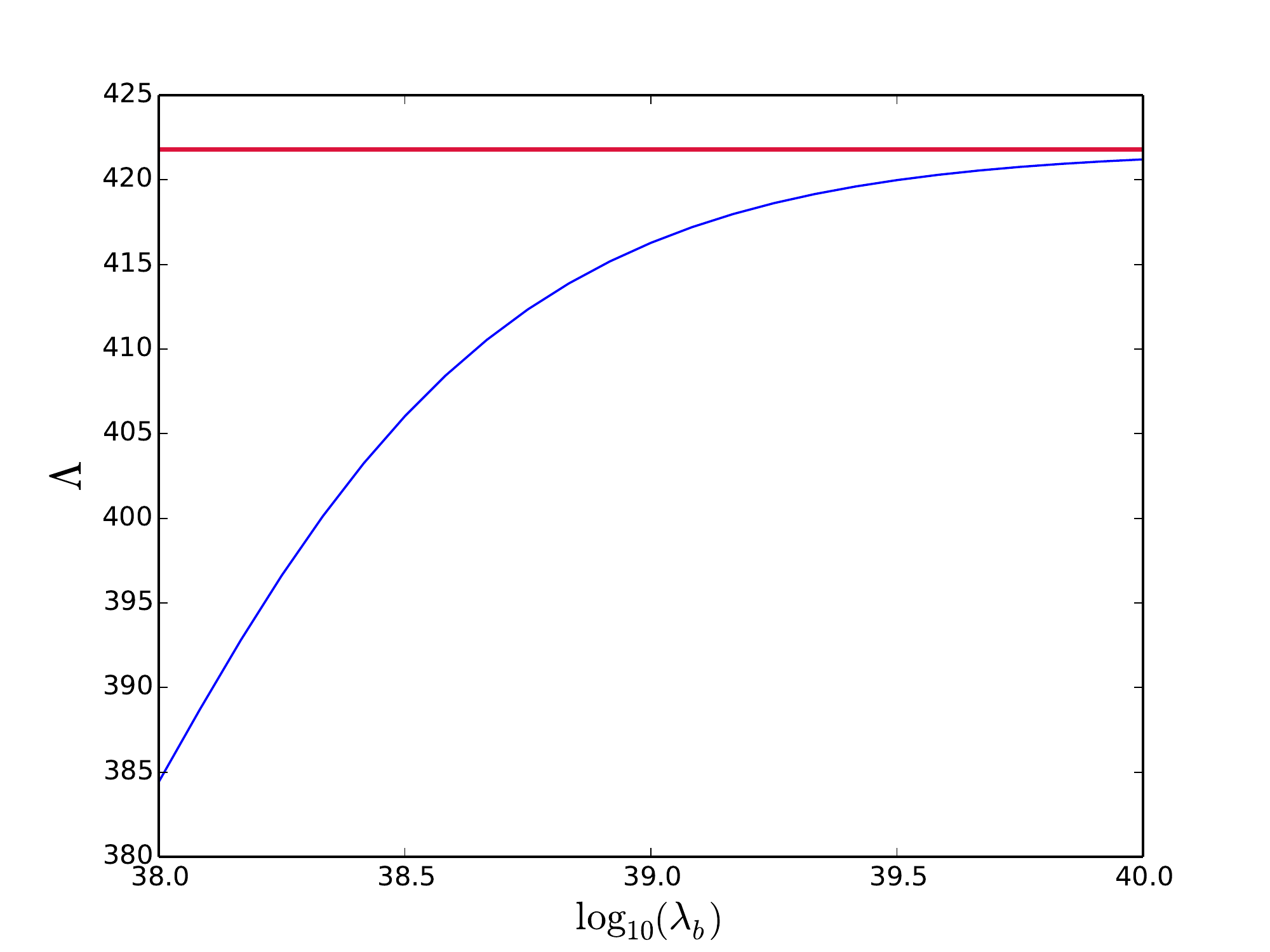}~~
\caption{The above figures present the variation of the dimensionless \tln\ $\Lambda$ with the brane tension $\lambda _{\rm b}$ from our numerical analysis for two choices of the equation of state of the neutron star. The curve on the left is for a tabulated equation-of-state parameter \cite{Alford:2004pf}, while that on the right is due to a polytropic equation of state with $\Gamma=5/3$. In both these figures, the red thick line depicts the asymptotic value of the dimensionless \tln\ $\Lambda$, which corresponds to the \gr\ limit. As is evident, the dimensionless \tln\ $\Lambda$ is non-zero and is smaller than the general relativistic value for a finite $\lambda_{\rm b}$, signalling the possible existence of extra dimensions. Note that the small-scale tests of gravitational interaction forbid one from numerically estimating of the brane tension $\lambda _{\rm b}$ below $10^{38}$ in units of mass density \cite{Germani:2001du}.}
\label{Lambdabrane}
\end{figure}
%%%%%%%%%%%%%%%%%%%%%%%%%%%%%
%%%%%%%%%%%%%%%%%%%%%%%%%%%%%
%%%%%%%%%%%%%%%%%%%%%%%%%%%%%
%%%%%%%%%%%%%%%%%%%%%%%%%%%%%

%%%%%%%%%%%%%%%%%%%%%%%%%%%%%
%%%%%%%%%%%%%%%%%%%%%%%%%%%%%
%%%%%%%%%%%%%%%%%%%%%%%%%%%%%
%%%%%%%%%%%%%%%%%%%%%%%%%%%%%
\begin{figure}[t!]
\includegraphics[scale=0.8]{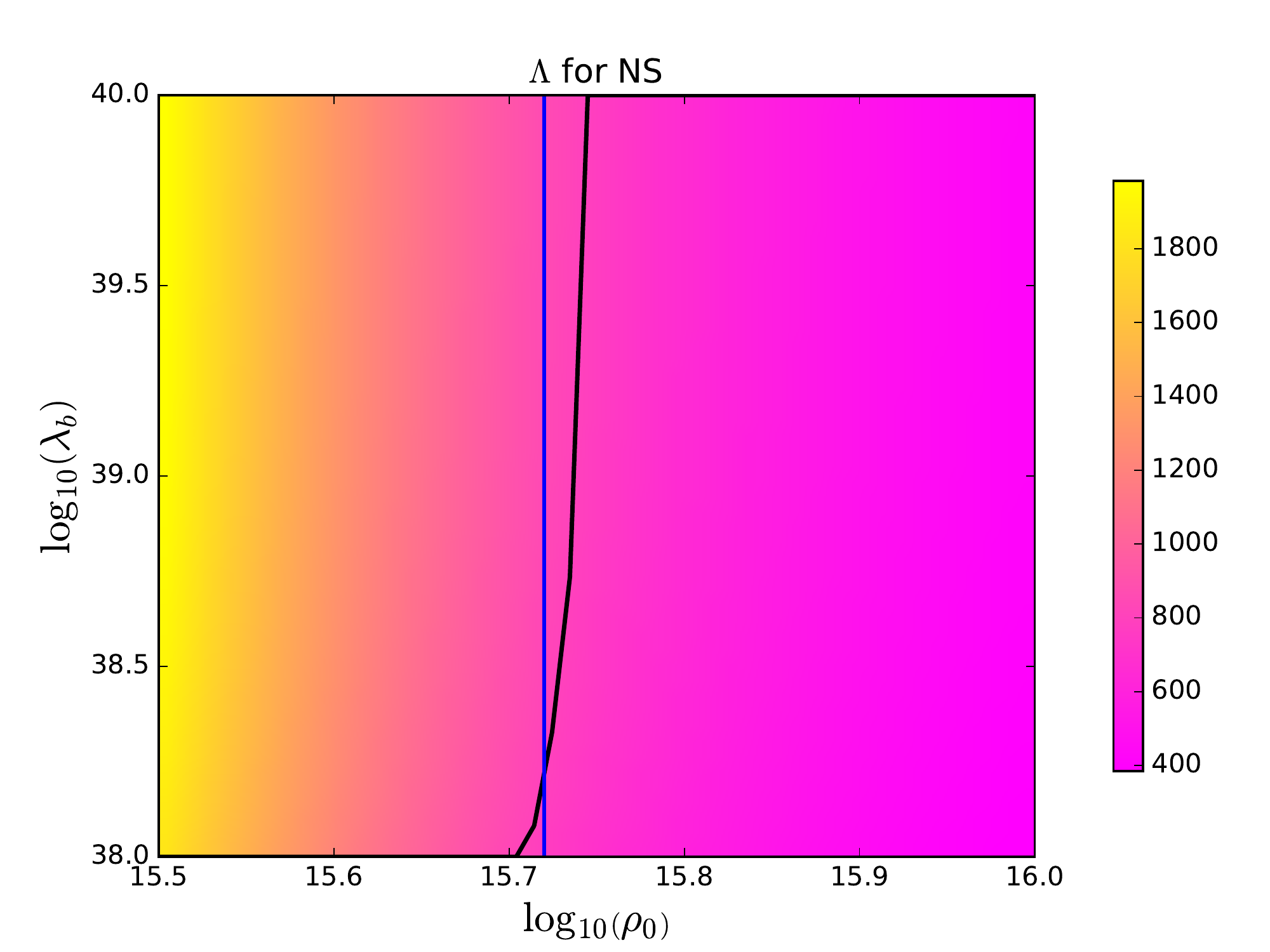}
\caption{The above figure depicts variation of the dimensionless \tln\ $\Lambda$ with the central density $\rho_{0}$ of a $\Gamma=5/3$ polytropic neutron star as well as the brane tension $\lambda _{\rm b}$, as computed from our numerical analysis. The \gr\ result is obtained by taking the limit $\lambda _{\rm b}\rightarrow \infty$, thus any finite value for $\lambda _{\rm b}$ will denote a departure from \gr. As evident from the panel on the right, $\Lambda$ is non-zero, and increases with decreasing central density, as it should. The black curve shows the $\Lambda=800$ line, such that points on the left are ruled out by the GW170817 event, while those on the right remain viable options \cite{TheLIGOScientific:2017qsa,Abbott:2018exr}. While the blue vertical line presents a certain central density, taken to be $10^{15.72}$ in $\textrm{gm}~\textrm{cm}^{-3}$. However, the value of $\Lambda$ changes with the central density and hence the $\Lambda=800$ curve being a mass dependent statement, provides a crude estimation of the bound from GW170817 event, which will suffice for our purpose. Moreover, the range of values for the brane tension $\lambda _{\rm b}$ is consistent with the small-scale tests of Newton's law.}
\label{tlnNS}
\end{figure}
%%%%%%%%%%%%%%%%%%%%%%%%%%%%%
%%%%%%%%%%%%%%%%%%%%%%%%%%%%%
%%%%%%%%%%%%%%%%%%%%%%%%%%%%%
%%%%%%%%%%%%%%%%%%%%%%%%%%%%%
The above analysis explicitly demonstrates that in the context of neutron stars there will be two sources of modifications to the \tln\ in presence of higher dimensions. The first such modification comes from the interior of the neutron star, as the master equation for the perturbation is different due to the presence of additional terms in the matter sector depending on the brane tension, along with a modified boundary condition at the surface of the star. This will lead to different values for the perturbation $H(r)$ and its derivative $H'(r)$ at the stellar surface. The second modification comes from the fact that the differential equation for gravitational perturbation in the exterior of the neutron star due to presence of Weyl stress will be different. Thus an estimation of the \tln\ with a modified differential equation and with modified boundary conditions will certainly differ from the general relativistic situation. 

Since in this particular context it is not possible to provide any analytical estimation for the \tln, or better the dimensionless \tln\ we have to resort to numerical methods. However, taking a cue from our earlier discussion regarding \tln\ for BHs, we can argue that possibly the effect of extra dimensions would be to decrease the numerical values for the dimensionless \tln\ from that in \gr. If this is indeed the case, there will be interesting consequences. For example, given a central density, the dimensionless \tln\ $\Lambda$ will be smaller in presence of higher dimensions i.e., in situations with finite $\lambda_{\rm b}$. Thus if GW experiments rule out some model with a given central density, they can again come into the picture if extra dimensions are taken into account. This may have observational ramifications as we will discuss below. 

Keeping these intriguing possibilities in mind we have performed a numerical analysis by solving the differential equation governing the behaviour of the even parity gravitational perturbation in the interior of the neutron star, which provided the numerical estimations for the gravitational perturbation and its derivative at the surface of the NS. These has later been used as the boundary conditions to solve for the gravitational perturbation in the exterior leading to a determination of the dimensionless \tln\ $\Lambda$. Of course, such estimations for the \tln\ will depend upon the choice of the EoS parameter for the material building up the neutron star. For illustration we have plotted these numerical estimates of $\Lambda$ against the brane tension $\lambda_{\rm b}$ for a tabulated EoS \cite{Alford:2004pf} and polytropic EoS in \ref{Lambdabrane}. Further numerical estimations for $\Lambda$ have also been presented for various choices of the brane tension $\lambda _{\rm b}$ and the central charge density of the neutron star in \ref{tlnNS}. As evident from \ref{tlnNS}, with an increase of the central density the dimensionless \tln\ decreases, since it becomes more difficult to deform the NS by the application of an external tidal field. Similarly, it is clear from \ref{tlnNS} (and also from \ref{Lambdabrane}) that the dimensionless \tln\ $\Lambda$ attains smaller and smaller values as the brane tension $\lambda _{\rm b}$ is decreased, i.e., as the system departs more and more from \gr. This being completely consistent with the theoretical consideration we had earlier. This explicitly demonstrates the consistency of the theoretical framework used in this work with the numerical analysis performed to get an estimation of the dimensionless \tln.  Following which in the next section we will discuss the implications for GW170817 and possible observability in future GW experiments. 
%%%%%%%%%%%%%%%%%%%%%%%%%%%%%%%%%%%%%%%%%%%%%%%%%%%%%%%%%%%%%%%%%%%%%%%%%%%%%%%%%%%%%%%%%%%%%%%%%%%
%%%%%%%%%%%%%%%%%%%%%%%%%%%%%%%%%%%%%%%%%%%%%%%%%%%%%%%%%%%%%%%%%%%%%%%%%%%%%%%%%%%%%%%%%%%%%%%%%%%
%%%%%%%%%%%%%%%%%%%%%%%%%%%%%%%%%%%%%%%%%%%%%%%%%%%%%%%%%%%%%%%%%%%%%%%%%%%%%%%%%%%%%%%%%%%%%%%%%%%
\section{Implications of GW170817 and future merger events}\label{imp_GW170817}

In this section we will comment on possible implications of the results derived above in the context of recent GW observation from NS-NS, e.g., GW170817 as well as BBH mergers. For GW170817, Advanced LIGO provides a constraint on the dimensionless \tln\ $\Lambda$, such that $\Lambda<800$ \cite{TheLIGOScientific:2017qsa}. This constraint in turn provide bounds on the parameter space of the EoS of the neutron star, by ruling out  several EoS leading to high central density of the neutron star. However, this analysis was completely within the context of \gr. As our numerical analysis depict, presence of extra dimensions reduce the estimations of dimensionless \tln\ $\Lambda$. Hence it follows that a significant portion of the EoS parameter space, earlier ruled out by \gr\ estimations will start becoming viable again. This has been clearly illustrated in \ref{tlnNS}. The quantity calculated is the dimensionless \tln\ $\Lambda$, for a polytropic EoS, with $\Gamma=5/3$, simultaneously as a function of the stellar central density and the brane tension $\lambda_{\rm b}$. The thick red line in \ref{tlnNS} presents the $\Lambda=800$ curve and as evident from the figure, the regions to the left of the $\Lambda=800$ curve (red thick line) are ruled out by the recent GW170817 measurement, while those on the right are still viable, assuming \gr. However, presence of extra dimension lead to a smaller value for $\Lambda$. Thus, given a central density (depicted by the blue vertical line), which may be ruled out in the context of \gr\ (corresponding to large-$\lambda_{\rm b}$ limit), may come into existence as one considers effect of extra spatial dimensions on the dimensionless \tln\ $\Lambda$. Further note that the parameter space for the brane tension considered in this work is completely consistent with the small-scale tests of Newton's law \cite{Long:2002wn,Hoyle:2004cw,Smullin:2005iv}. Thus in future observations, if the \tln\ gets more and more accurately measured by the advanced GW detectors, that in turn will provide stringent bounds on the brane tension $\lambda _{\rm b}$, which will be more accurate compared to the small-scale tests of Newton's law and hence we will really be probing some of the microscopic features of spacetime through GW experiments. This is a direct consequence of the presence of extra dimensions on the NS-NS merger. 

For observations related to BBH mergers, on the other hand, the situation is more subtle, but also more predictable. The dimensionless \tln\ $\Lambda$ values associated with BHs in presence of extra dimension are computed numerically by asymptotic matching of the other dimensionless \tln\ $k_2$. As evident from \ref{Lambdabh}, numerical estimates for the dimensionless \tln\ $\Lambda$ are in the range of $\sim -10$ to $-20$. Further from \ref{Lambdabh}, we can also see the variation of the dimensionless \tln\ $\Lambda$ for brane world BH with the parameter $\beta$, which is also not very large $\sim 20\%$. Given the present sensitivity of Advanced LIGO detectors, and their associated errors, a negative $\Lambda$ value of a few hundreds would have been easily detected, but that of few tens (as is the case here) will probably be lost in noise. On a brighter note, though, the upcoming Einstein Telescope (or LISA) will increase the sensitivity by an order of magnitude, meaning that lower values for the dimensionless \tln\ $\Lambda$ could then be detected, if they turned out to be negative, that may act as a very good testbed for higher dimensional theories. 
%%%%%%%%%%%%%%%%%%%%%%%%%%%%%%%%%%%%%%%%%%%%%%%%%%%%%%%%%%%%%%%%%%%%%%%%%%%%%%%%%%%%%%%%%%%%%%%%%%%
%%%%%%%%%%%%%%%%%%%%%%%%%%%%%%%%%%%%%%%%%%%%%%%%%%%%%%%%%%%%%%%%%%%%%%%%%%%%%%%%%%%%%%%%%%%%%%%%%%%
%%%%%%%%%%%%%%%%%%%%%%%%%%%%%%%%%%%%%%%%%%%%%%%%%%%%%%%%%%%%%%%%%%%%%%%%%%%%%%%%%%%%%%%%%%%%%%%%%%%
\section{Conclusions}

Understanding possible implications of theories beyond \gr\ has become a topic of significant importance in the recent years, thanks to the detection of GWs from binary BHs as well as binary NSs providing a first hand experience to the strong gravity regime. Among various other possibilities existence of extra dimensions and its implications and observability in the context of GWs are of significant interest. Following this trend, in this work we explored the effect of higher spatial dimensions on the \tln\ of BHs and NSs. For this purpose, we have started with an understanding of the modifications to the static, even parity gravitational perturbation in the exterior of a NS or BH, pertaining to the existence of extra dimensions. These modifications to the differential equation satisfied by gravitational perturbation will also lead to possible modifications to the tidal Love number. 

In particular, we have explicitly demonstrated, using both theoretical as well as numerical techniques that the presence of an extra dimension will, beyond doubt, make tidal Love numbers non-zero and more importantly \emph{negative} for brane world BHs. On the other hand it is well known that for BHs in \gr\ the \tln\ vanishes. Incidentally the \gr\ result can also be derived by taking appropriate limit of our higher dimensional result. We would like to emphasize that even though the idea of non-zero tidal Love numbers for BHs are not new, but specifically for the case of extra dimensions, we find the tidal Love numbers to be negative, unlike other scenarios. The negativity of \tln\ for BHs is an additional distinguishing feature of the extra dimension, and possibly can be exploited using real data in future for further confirmation. 

We have further demonstrated the relevant modifications to the differential equations governing gravitational perturbations in the interior of the neutron star as well. In general this will lead to a two fluid system with the Weyl fluid, inherited from higher dimension being one, while the fluid inside the neutron star being another. However due to its difficult nature we have considered a simpler situation in which the Weyl fluid is non-existent inside the neutron star but is certainly present outside, still preserving continuity of physical quantities on the surface of the NS. In this particular context the essential modifications done to the equations satisfied by the gravitational perturbations in the interior of the neutron star involves the term $\rho^{2}/\lambda _{\rm b}$, where $\rho$ is the energy density of the fluid filling the interior of the NS and $\lambda _{\rm b}$ is the brane tension. As $\lambda _{\rm b}\rightarrow \infty$ the \gr\ result can be obtained. As we have explicitly demonstrated by solving these equations numerically that in the case of NS-NS or NS-BH binaries, the presence of extra dimensions will induce deviations in dimensionless \tln\ $\Lambda$ from its {\it{expected}} general relativistic behaviour, which essentially decreases the numerical estimations of the \tln. This results in an interesting possibility, namely a certain parameter space associated with central density of the NS  (which was earlier ruled out by GW170817 using general-relativistic methods) may be viable for a finite $\lambda _{\rm b}$, signalling possible existence of extra dimensions. Moreover, future GW observations will constrain the dimensionless \tln\ $\Lambda$ to a greater accuracy, which in turn may lead to further stringent constraint, better than the existing ones, on the brane tension $\lambda _{\rm b}$. Further, given these modifications to the dimensionless \tln\ $\Lambda$, there may be a very good chance of detection in future GW detectors because of the low Signal-to-Noise-Ratio of these measurements. Thus we have explicitly demonstrated using both theoretical and numerical considerations a modification to the dimensionless \tln\ $\Lambda$ (or, $k_{2}$) for a specific scenario of compactified extra spatial dimensions. The techniques employed here are quite general, and can equivalently be applied to understand the nature of \tln\ in various other theories of gravity beyond \gr\ as well. We leave these questions for future studies.
%%%%%%%%%%%%%%%%%%%%%%%%%%%%%%%%%%%%%%%%%%%%%%%%%%%%%%%%%%%%%%%%%%%%%%%%%%%%%%%%%%%%%%%%%%%%%%%%%%%
%%%%%%%%%%%%%%%%%%%%%%%%%%%%%%%%%%%%%%%%%%%%%%%%%%%%%%%%%%%%%%%%%%%%%%%%%%%%%%%%%%%%%%%%%%%%%%%%%%%
%%%%%%%%%%%%%%%%%%%%%%%%%%%%%%%%%%%%%%%%%%%%%%%%%%%%%%%%%%%%%%%%%%%%%%%%%%%%%%%%%%%%%%%%%%%%%%%%%%%
\section*{Acknowledgements}

We thank Nils Andersson and Alessandro Nagar for helpful discussions and crucial insights. We also thank Bhooshan Gadre for carefully reading the manuscript and making useful suggestions. Research of SC is supported by INSPIRE Faculty Fellowship (Reg. No. DST/INSPIRE/04/2018/000893) and he also thanks IUCAA, Pune for warm hospitality where a part of this work was done. Research of SSG is partially supported by the SERB-Extra Mural Research grant (EMR/2017/001372), Government of India. This work was supported in part by NSF Grant PHY-1506497 and the Navajbai Ratan Tata Trust. This work has been assigned the LIGO document number: LIGO-P1800288.
%%%%%%%%%%%%%%%%%%%%%%%%%%%%%%%%%%%%%%%%%%%%%%%%%%%%%%%%%%%%%%%%%%%%%%%%%%%%%%%%%%%%%%%%%%%%%%%%%%%
%%%%%%%%%%%%%%%%%%%%%%%%%%%%%%%%%%%%%%%%%%%%%%%%%%%%%%%%%%%%%%%%%%%%%%%%%%%%%%%%%%%%%%%%%%%%%%%%%%%
%%%%%%%%%%%%%%%%%%%%%%%%%%%%%%%%%%%%%%%%%%%%%%%%%%%%%%%%%%%%%%%%%%%%%%%%%%%%%%%%%%%%%%%%%%%%%%%%%%%
\appendix

\labelformat{section}{Appendix #1}
\labelformat{subsection}{Appendix #1}
\labelformat{subsubsection}{Appendix #1}
%%%%%%%%%%%%%%%%%%%%%%%%%%%%%%%%%%%%%%%%%%%%%%%%%%%%%%%%%%%%%%%%%%%%%%%%%%%%%%%%%%%%%%%%%%%%%%%%%%%
%%%%%%%%%%%%%%%%%%%%%%%%%%%%%%%%%%%%%%%%%%%%%%%%%%%%%%%%%%%%%%%%%%%%%%%%%%%%%%%%%%%%%%%%%%%%%%%%%%%
%%%%%%%%%%%%%%%%%%%%%%%%%%%%%%%%%%%%%%%%%%%%%%%%%%%%%%%%%%%%%%%%%%%%%%%%%%%%%%%%%%%%%%%%%%%%%%%%%%%
\section*{Appendices}
%%%%%%%%%%%%%%%%%%%%%%%%%%%%%%%%%%%%%%%%%%%%%%%%%%%%%%%%%%%%%%%%%%%%%%%%%%%%%%%%%%%%%%%%%%%%%%%%%%%
%%%%%%%%%%%%%%%%%%%%%%%%%%%%%%%%%%%%%%%%%%%%%%%%%%%%%%%%%%%%%%%%%%%%%%%%%%%%%%%%%%%%%%%%%%%%%%%%%%%
%%%%%%%%%%%%%%%%%%%%%%%%%%%%%%%%%%%%%%%%%%%%%%%%%%%%%%%%%%%%%%%%%%%%%%%%%%%%%%%%%%%%%%%%%%%%%%%%%%%
\section{Solving the differential equation for even parity gravitational perturbation}\label{TLN_App_v2_Diff}

Let us try to solve for \ref{TLN_v2_41} and for that purpose, it will be helpful to define a new quantity $\bar{\beta}^{2}=4\{1+(\beta/2)\}\{1+\beta\}^{-1}$. Thus the differential equation takes the following structure,
%%%%%%%%%%%%%%%%%%%%%%%%%%%%%%%%%%%%%%%%%%%%%%%%%%%%%%%%%%%%%%
\begin{align}\label{TLN_v1_App_01}
\left\{y^{2}-1\right\}\partial _{y}^{2}H+2y\partial _{y}H
-\left\{\ell\left(\ell+1\right)+\frac{\bar{\beta}^{2}}{y^{2}-1}\right\}H=0
\end{align}
%%%%%%%%%%%%%%%%%%%%%%%%%%%%%%%%%%%%%%%%%%%%%%%%%%%%%%%%%%%%%%
This differential equation is exactly the same as the differential equation satisfied by the Associated Lengendre polynomials $P^{\ell}_{m}(y)$ and $Q^{\ell}_{m}(y)$, but with the exception of $m=\bar{\beta}$ now being fractional. However in the context of \gr\  $\beta=0$ and we have the usual case with $m=2$. The solution of this equation, found from \cite{Abramowitz:Book} is given by:
%%%%%%%%%%%%%%%%%%%%%%%%%%%%%%%%%%%%%%%%%%%%%%%%%%%%%%%%%%%%%%
\begin{equation}\label{TLN_v1_App_02}
H(\ell,m;y) = A_{1} P_{\ell}^m(y)+B_{1} Q_{\ell}^m(y)
\end{equation}
%%%%%%%%%%%%%%%%%%%%%%%%%%%%%%%%%%%%%%%%%%%%%%%%%%%%%%%%%%%%%%
where $P_{\ell}^m(y)$ and $Q_{\ell}^m(y)$ are the Legendre functions of the first and the second kind respectively. Further the arbitrary constants appearing from solving the above second order differential equations are $A_{1}$ and $B_{1}$, respectively. Since we are interested in the asymptotic behaviour we would like to write down the above functions as polynomials in $y$. This can be achieved by first expanding them in terms of Hypergeometric functions and then employing a power series expansion. When written in terms of the confluent Hypergeometric functions $_{2}F_{1}$ we obtain,
%%%%%%%%%%%%%%%%%%%%%%%%%%%%%%%%%%%%%%%%%%%%%%%%%%%%%%%%%%%%%%
\begin{align}
Q_{\ell}^m(y)&= e^{im\pi} 2^{-\ell-1} \sqrt{\pi}\frac{\Gamma(\ell+m+1)}{\Gamma(\ell+\frac{3}{2})} y^{-\ell-m-1}(y^2-1)^{m/2}
~_{2}F_{1} \left(1+\frac{\ell+m}{2},\frac{1+\ell+m}{2},\ell+\frac{3}{2};\frac{1}{y^2}\right) 
\label{TLN_v1_App_03}
\\
P_{\ell}^m(y)&=2^{-\ell-1} \pi^{-1/2}\frac{\Gamma(-\frac{1}{2}-\ell)}{\Gamma(-\ell-m)} y^{-\ell+m-1)}(y^2-1)^{-m/2}
~_{2}F_{1}\left(\frac{1}{2}+\frac{\ell-m}{2},1+\frac{\ell-m}{2},\ell+\frac{3}{2};\frac{1}{y^2}\right)
\nonumber
\\
&+2^{\ell}\frac{\Gamma(\frac{1}{2}+\ell)}{\Gamma(1+\ell-m)}y^{\ell+m}\left(y^{2}-1\right)^{-m/2}
~_{2}F_{1}\left(-\frac{\ell+m}{2},\frac{1}{2}-\frac{\ell+m}{2},\frac{1}{2}-\ell;\frac{1}{y^{2}}\right)
\label{TLN_v1_App_04}
\end{align} 
%%%%%%%%%%%%%%%%%%%%%%%%%%%%%%%%%%%%%%%%%%%%%%%%%%%%%%%%%%%%%%
For argument smaller than unity, the hypergeometric functions have the following polynomial expansion,
%%%%%%%%%%%%%%%%%%%%%%%%%%%%%%%%%%%%%%%%%%%%%%%%%%%%%%%%%%%%%%
\begin{align}\label{TLN_v1_App_05}
_{2}F_{1}(a,b,c;x)=\frac{\Gamma(c)}{\Gamma(a)\Gamma(b)}\sum _{j=0}^{\infty}\frac{\Gamma(a+j)\Gamma(b+j)}{\Gamma(c+j)}\frac{x^{j}}{j!}
\end{align}
%%%%%%%%%%%%%%%%%%%%%%%%%%%%%%%%%%%%%%%%%%%%%%%%%%%%%%%%%%%%%%
Using this expression for the confluent hypergeometric series, we can rewrite the associated Legendre polynomial of second kind, namely $Q^{m}_{\ell}$ from \ref{TLN_v1_App_03} as,
%%%%%%%%%%%%%%%%%%%%%%%%%%%%%%%%%%%%%%%%%%%%%%%%%%%%%%%%%%%%%%
\begin{align}\label{TLN_v1_App_06}
Q_{\ell}^m(y)&= e^{im\pi} 2^{-\ell-1} \sqrt{\pi}\frac{\Gamma(\ell+m+1)}{\Gamma(\ell+\frac{3}{2})} y^{-\ell-m-1}(y^2-1)^{m/2}
\nonumber
\\
&~\times \frac{\Gamma(\ell+\frac{3}{2})}{\Gamma(1+\frac{\ell+m}{2})\Gamma(\frac{1+\ell+m}{2})}\sum _{j=0}^{\infty}\frac{\Gamma(\frac{1+\ell+m}{2}+j)\Gamma(1+\frac{\ell+m}{2}+j)}{\Gamma(\ell+\frac{3}{2}+j)}\frac{y^{-2j}}{j!}
\end{align} 
%%%%%%%%%%%%%%%%%%%%%%%%%%%%%%%%%%%%%%%%%%%%%%%%%%%%%%%%%%%%%%
It will be worthwhile to define,
%%%%%%%%%%%%%%%%%%%%%%%%%%%%%%%%%%%%%%%%%%%%%%%%%%%%%%%%%%%%%%
\begin{align}
\alpha(\ell,m)&= e^{im\pi} 2^{-\ell-1} \sqrt{\pi}\frac{\Gamma(\ell+m+1)}{\Gamma(1+\frac{\ell+m}{2})\Gamma(\frac{1+\ell+m}{2})} 
\label{TLN_v1_App_07a}
\\
\beta_{j}(\ell,m)&=\frac{\Gamma(\frac{1+\ell+m}{2}+j)\Gamma(1+\frac{\ell+m}{2}+j)}{\Gamma(\ell+\frac{3}{2}+j)}
\label{TLN_v1_App_07b}
\end{align} 
%%%%%%%%%%%%%%%%%%%%%%%%%%%%%%%%%%%%%%%%%%%%%%%%%%%%%%%%%%%%%%
in terms of which $Q^{m}_{\ell}(y)$ has the following expression,
%%%%%%%%%%%%%%%%%%%%%%%%%%%%%%%%%%%%%%%%%%%%%%%%%%%%%%%%%%%%%%
\begin{align}\label{TLN_v1_App_08}
Q_{\ell}^m(y)&= \alpha(\ell,m)y^{-\ell-m-1}(y^2-1)^{m/2}\sum _{n=0}^{\infty}\beta_{n}(\ell,m)\frac{y^{-2n}}{n!}
\end{align} 
%%%%%%%%%%%%%%%%%%%%%%%%%%%%%%%%%%%%%%%%%%%%%%%%%%%%%%%%%%%%%%
Since we are interested in the asymptotic limit, we can expand the above in powers of $y^{-1}$, immediately leading to,
%%%%%%%%%%%%%%%%%%%%%%%%%%%%%%%%%%%%%%%%%%%%%%%%%%%%%%%%%%%%%%
\begin{align}\label{TLN_v1_App_09}
Q_{\ell}^m(y)&= \alpha(\ell,m)y^{-\ell-1}(1-y^{-2})^{m/2}\sum _{j=0}^{\infty}\beta_{j}(\ell,m)\frac{y^{-2j}}{j!}
\nonumber
\\
&=\alpha(\ell,m)y^{-\ell-1}\left\lbrace \sum _{k=0}^{\infty}\left(-1\right)^{k}\frac{\left(\frac{m}{2}\right)!}{k!\left(\frac{m}{2}-k\right)!}y^{-2k}\right\rbrace 
\left \lbrace\sum _{j=0}^{\infty}\beta_{j}(\ell,m)\frac{y^{-2j}}{j!}\right\rbrace
\end{align} 
%%%%%%%%%%%%%%%%%%%%%%%%%%%%%%%%%%%%%%%%%%%%%%%%%%%%%%%%%%%%%%
By expanding this series we obtain the first three nontrivial terms in the expression for $Q^{m}_{\ell}$:
%%%%%%%%%%%%%%%%%%%%%%%%%%%%%%%%%%%%%%%%%%%%%%%%%%%%%%%%%%%%%%
\begin{align}\label{TLN_v1_App_10}
Q_{\ell}^m(y)\simeq \alpha(\ell,m)y^{-\ell-1}\Bigg[\beta _{0}(\ell,m)
&+\left\lbrace -\left(\frac{m}{2}\right)\beta _{0}(\ell,m)+\beta _{1}(\ell,m)\right\rbrace y^{-2}
\nonumber
\\
&+\left\lbrace \frac{1}{2}\left(\frac{m}{2}\right)\left(\frac{m}{2}-1\right)\beta _{0}(\ell,m)
-\left(\frac{m}{2}\right)\beta _{1}(\ell,m)+\frac{1}{2}\beta _{2}(\ell,m)\right\rbrace y^{-4}\Bigg]
\end{align} 
%%%%%%%%%%%%%%%%%%%%%%%%%%%%%%%%%%%%%%%%%%%%%%%%%%%%%%%%%%%%%%
Let us concentrate on the associated Legendre polynomial of first kind $P^{m}_{\ell}$, given by \ref{TLN_v1_App_04}. Using the series expansion of the hypergeometric function, we obtain,
%%%%%%%%%%%%%%%%%%%%%%%%%%%%%%%%%%%%%%%%%%%%%%%%%%%%%%%%%%%%%%
\begin{align}\label{TLN_v1_App_11}
P_{\ell}^m(y)&=2^{-\ell-1} \pi^{-1/2}\frac{\Gamma(-\frac{1}{2}-\ell)}{\Gamma(-\ell-m)} y^{-\ell+m-1}(y^2-1)^{-m/2}
\nonumber
\\
&\qquad \qquad \times \frac{\Gamma(\ell+\frac{3}{2})}{\Gamma(\frac{1}{2}+\frac{\ell-m}{2})\Gamma(1+\frac{\ell-m}{2})}
\sum _{j=0}^{\infty}\frac{\Gamma(1+\frac{\ell-m}{2}+j)\Gamma(\frac{1}{2}+\frac{\ell-m}{2}+j)}{\Gamma(\ell+\frac{3}{2}+j)}
\frac{y^{-2j}}{j!}
\nonumber
\\
&+2^{\ell}\frac{\Gamma(\frac{1}{2}+\ell)}{\Gamma(1+\ell-m)}y^{\ell+m}\left(y^{2}-1\right)^{-m/2}
\nonumber
\\
&\qquad \qquad \times \frac{\Gamma(\frac{1}{2}-\ell)}{\Gamma(-\frac{\ell+m}{2})\Gamma(\frac{1}{2}-\frac{\ell+m}{2})}\sum _{j=0}^{\infty}\frac{\Gamma(-\frac{\ell+m}{2}+j)\Gamma(\frac{1}{2}-\frac{\ell+m}{2}+j)}{\Gamma(\frac{1}{2}-\ell+j)}\frac{y^{-2j}}{j!}
\end{align} 
%%%%%%%%%%%%%%%%%%%%%%%%%%%%%%%%%%%%%%%%%%%%%%%%%%%%%%%%%%%%%%
For convenience, in this case as well, we will introduce four quantities,
%%%%%%%%%%%%%%%%%%%%%%%%%%%%%%%%%%%%%%%%%%%%%%%%%%%%%%%%%%%%%%
\begin{align}
\gamma(\ell,m)&=2^{-\ell-1} \pi^{-1/2}\frac{\Gamma(-\frac{1}{2}-\ell)}{\Gamma(-\ell-m)} 
\frac{\Gamma(\ell+\frac{3}{2})}{\Gamma(\frac{1}{2}+\frac{\ell-m}{2})\Gamma(1+\frac{\ell-m}{2})}
\label{TLN_v1_App_12a}
\\
\sigma _{j}(\ell,m)&=\frac{\Gamma(1+\frac{\ell-m}{2}+j)\Gamma(\frac{1}{2}+\frac{\ell-m}{2}+j)}{\Gamma(\ell+\frac{3}{2}+j)}
\label{TLN_v1_App_12b}
\\
\chi(\ell,m)&=2^{\ell}\frac{\Gamma(\frac{1}{2}+\ell)}{\Gamma(1+\ell-m)}
\frac{\Gamma(\frac{1}{2}-\ell)}{\Gamma(-\frac{\ell+m}{2})\Gamma(\frac{1}{2}-\frac{\ell+m}{2})}
\label{TLN_v1_App_12c}
\\
\Pi_{j}(\ell,m)&=\frac{\Gamma(-\frac{\ell+m}{2}+j)\Gamma(\frac{1}{2}-\frac{\ell+m}{2}+j)}{\Gamma(\frac{1}{2}-\ell+j)}
\label{TLN_v1_App_12d}
\end{align} 
%%%%%%%%%%%%%%%%%%%%%%%%%%%%%%%%%%%%%%%%%%%%%%%%%%%%%%%%%%%%%%
and hence the associated Legendre polynomial can be written as,
%%%%%%%%%%%%%%%%%%%%%%%%%%%%%%%%%%%%%%%%%%%%%%%%%%%%%%%%%%%%%%
\begin{align}\label{TLN_v1_App_13}
P_{\ell}^m(y)&=\gamma(\ell,m)y^{-\ell+m-1}(y^2-1)^{-m/2}
\sum _{j=0}^{\infty}\sigma _{j}(\ell,m)\frac{y^{-2j}}{j!}
\nonumber
\\
&+\chi(\ell,m)y^{\ell+m}\left(y^{2}-1\right)^{-m/2}
\sum _{j=0}^{\infty}\Pi_{j}(\ell,m)\frac{y^{-2j}}{j!}
\end{align} 
%%%%%%%%%%%%%%%%%%%%%%%%%%%%%%%%%%%%%%%%%%%%%%%%%%%%%%%%%%%%%%
Expanding the above expression for large values of $y$, we obtain,
%%%%%%%%%%%%%%%%%%%%%%%%%%%%%%%%%%%%%%%%%%%%%%%%%%%%%%%%%%%%%%
\begin{align}\label{TLN_v1_App_14}
P_{\ell}^m(y)&=\gamma(\ell,m)y^{-\ell-1}(1-y^{-2})^{-m/2}\sum _{j=0}^{\infty}\sigma _{j}(\ell,m)\frac{y^{-2j}}{j!}
+\chi(\ell,m)y^{\ell}\left(1-y^{-2}\right)^{-m/2}\sum _{j=0}^{\infty}\Pi_{j}(\ell,m)\frac{y^{-2j}}{j!}
\nonumber
\\
&=\gamma(\ell,m)y^{-\ell-1}\sum_{k=0}^{\infty}\frac{\left(\frac{m}{2}+k-1\right)!}{k!\left(\frac{m}{2}-1\right)!}y^{-2k}\sum _{j=0}^{\infty}\sigma _{j}(\ell,m)\frac{y^{-2j}}{j!}
\nonumber
\\
&\qquad \qquad +\chi(\ell,m)y^{\ell}\sum_{k=0}^{\infty}\frac{\left(\frac{m}{2}+k-1\right)!}{k!\left(\frac{m}{2}-1\right)!}y^{-2k}\sum _{j=0}^{\infty}\Pi_{j}(\ell,m)\frac{y^{-2j}}{j!}
\end{align} 
%%%%%%%%%%%%%%%%%%%%%%%%%%%%%%%%%%%%%%%%%%%%%%%%%%%%%%%%%%%%%%
Keeping the first few non-trivial terms, the above expression yields,
%%%%%%%%%%%%%%%%%%%%%%%%%%%%%%%%%%%%%%%%%%%%%%%%%%%%%%%%%%%%%%
\begin{align}\label{TLN_v1_App_15}
P_{\ell}^m(y)\simeq \gamma(\ell,m)y^{-\ell-1}\Bigg[\sigma _{0}(\ell,m)
&+\left\lbrace \left(\frac{m}{2}\right)\sigma _{0}(\ell,m)+\sigma _{1}(\ell,m)\right\rbrace y^{-2}
\nonumber
\\
&+\left\lbrace \frac{1}{2}\left(\frac{m}{2}\right)\left(\frac{m}{2}+1\right)\sigma _{0}(\ell,m)
+\left(\frac{m}{2}\right)\sigma _{1}(\ell,m)+\frac{1}{2}\sigma _{2}(\ell,m)\right\rbrace y^{-4}\Bigg]
\nonumber
\\
+\chi(\ell,m)y^{\ell}\Bigg[\Pi _{0}(\ell,m)
&+\left\lbrace \left(\frac{m}{2}\right)\Pi _{0}(\ell,m)+\Pi _{1}(\ell,m)\right\rbrace y^{-2}
\nonumber
\\
&+\left\lbrace \frac{1}{2}\left(\frac{m}{2}\right)\left(\frac{m}{2}+1\right)\Pi _{0}(\ell,m)
+\left(\frac{m}{2}\right)\Pi _{1}(\ell,m)+\frac{1}{2}\Pi _{2}(\ell,m)\right\rbrace y^{-4}\Bigg]
\end{align} 
%%%%%%%%%%%%%%%%%%%%%%%%%%%%%%%%%%%%%%%%%%%%%%%%%%%%%%%%%%%%%%
This completes our discussion of the general solution and first few non-trivial terms in its asymptotic expansion. 

However, keeping the future applications in mind, let us concentrate on the $\ell=2$ situation and the leading order contribution, which yields,
%%%%%%%%%%%%%%%%%%%%%%%%%%%%%%%%%%%%%%%%%%%%%%%%%%%%%%%%%%%%%%
\begin{align}\label{TLN_v1_App_16}
H(2,m;y) &\simeq A_{1}\left\lbrace \gamma(2,m)\sigma _{0}(2,m)y^{-3}+\chi(2,m)\Pi _{0}(2,m)y^{2} \right\rbrace 
+B_{1} \alpha(2,m)\beta _{0}(2,m)y^{-3}
\nonumber
\\
&=\left\lbrace \gamma(2,m)\sigma _{0}(2,m)A_{1}+\alpha(2,m)\beta _{0}(2,m)B_{1} \right\rbrace y^{-3}
+\chi(2,m)\Pi _{0}(2,m)A_{1}y^{2}
\end{align}
%%%%%%%%%%%%%%%%%%%%%%%%%%%%%%%%%%%%%%%%%%%%%%%%%%%%%%%%%%%%%%
Given the results in \ref{TLN_v1_App_07a}, \ref{TLN_v1_App_07b} as well as \ref{TLN_v1_App_12a}, \ref{TLN_v1_App_12b}, \ref{TLN_v1_App_12c} and \ref{TLN_v1_App_12d} we finally obtain, 
%%%%%%%%%%%%%%%%%%%%%%%%%%%%%%%%%%%%%%%%%%%%%%%%%%%%%%%%%%%%%%
\begin{align}
\alpha(2,m)&=\frac{\sqrt{\pi}}{8}e^{im\pi}\frac{\Gamma(3+m)}{\Gamma(2+\frac{m}{2})\Gamma(\frac{3}{2}+\frac{m}{2})};\qquad
\beta_{0}(2,m)=\frac{\Gamma(2+\frac{m}{2})\Gamma(\frac{3}{2}+\frac{m}{2})}{\Gamma(\frac{7}{2})}
\label{TLN_v1_App_17a}
\\
\gamma(2,m)&=\frac{1}{8\sqrt{\pi}}\frac{\Gamma(-\frac{5}{2})}{\Gamma(-2-m)}\frac{\Gamma(\frac{7}{2})}{\Gamma(\frac{3}{2}-\frac{m}{2})\Gamma(2-\frac{m}{2})};\qquad
\sigma _{0}(2,m)=\frac{\Gamma(\frac{3}{2}-\frac{m}{2})\Gamma(2-\frac{m}{2})}{\Gamma(\frac{7}{2})}
\label{TLN_v1_App_17b}
\\
\chi(2,m)&=4\frac{\Gamma(\frac{5}{2})}{\Gamma(3-m)}\frac{\Gamma(-\frac{3}{2})}{\Gamma(-1-\frac{m}{2})\Gamma(-\frac{1}{2}-\frac{m}{2})};\qquad 
\Pi_{0}(2,m)=\frac{\Gamma(-1-\frac{m}{2})\Gamma(-\frac{1}{2}-\frac{m}{2})}{\Gamma(-\frac{3}{2})}
\label{TLN_v1_App_17c}
\end{align}
%%%%%%%%%%%%%%%%%%%%%%%%%%%%%%%%%%%%%%%%%%%%%%%%%%%%%%%%%%%%%%
Using the above results we finally obtain,
%%%%%%%%%%%%%%%%%%%%%%%%%%%%%%%%%%%%%%%%%%%%%%%%%%%%%%%%%%%%%%
\begin{align}\label{TLN_v1_App_18}
\alpha(2,m)\beta_{0}(2,m)=\frac{\Gamma(3+m)}{15}e^{im\pi};\quad 
\gamma(2,m)\sigma _{0}(2,m)=\frac{1}{8\sqrt{\pi}}\frac{\Gamma(-\frac{5}{2})}{\Gamma(-2-m)};\quad
\chi(2,m)\Pi_{0}(2,m)=\frac{4\Gamma(\frac{5}{2})}{\Gamma(3-m)}
\end{align}
%%%%%%%%%%%%%%%%%%%%%%%%%%%%%%%%%%%%%%%%%%%%%%%%%%%%%%%%%%%%%%
Further the following result: $\Gamma(n+1)=n\Gamma(n)$ can be used, to obtain
%%%%%%%%%%%%%%%%%%%%%%%%%%%%%%%%%%%%%%%%%%%%%%%%%%%%%%%%%%%%%%
\begin{align}\label{TLN_v1_App_19}
\Gamma(-\frac{5}{2})=\left(-\frac{2}{5}\right)\Gamma(-\frac{3}{2})=\left(-\frac{2}{5}\right)\left(-\frac{2}{3}\right)\Gamma(-\frac{1}{2})
=-\frac{8\sqrt{\pi}}{15}
\end{align}
%%%%%%%%%%%%%%%%%%%%%%%%%%%%%%%%%%%%%%%%%%%%%%%%%%%%%%%%%%%%%%
as well as, $\Gamma(5/2)=(3/4)\sqrt{\pi}$. Thus the solution becomes,
%%%%%%%%%%%%%%%%%%%%%%%%%%%%%%%%%%%%%%%%%%%%%%%%%%%%%%%%%%%%%%
\begin{align}\label{TLN_v1_App_20}
H(2,m;y)=\left\lbrace \frac{3A_{1}\sqrt{\pi}}{\Gamma(3-m)} \right\rbrace y^{2}
+\left\lbrace -\frac{A_{1}}{15\Gamma(-2-m)} 
+\frac{B_{1}\Gamma(3+m) e^{i\pi \bar{\beta}}}{15} \right\rbrace y^{-3}
\end{align}
%%%%%%%%%%%%%%%%%%%%%%%%%%%%%%%%%%%%%%%%%%%%%%%%%%%%%%%%%%%%%%
This is the result we have used in the main text.
%%%%%%%%%%%%%%%%%%%%%%%%%%%%%%%%%%%%%%%%%%%%%%%%%%%%%%%%%%%%%%%%%%%%%%%%%%%%%%%%%%%%%%%%%%%%%%%%%%%
%%%%%%%%%%%%%%%%%%%%%%%%%%%%%%%%%%%%%%%%%%%%%%%%%%%%%%%%%%%%%%%%%%%%%%%%%%%%%%%%%%%%%%%%%%%%%%%%%%%
%%%%%%%%%%%%%%%%%%%%%%%%%%%%%%%%%%%%%%%%%%%%%%%%%%%%%%%%%%%%%%%%%%%%%%%%%%%%%%%%%%%%%%%%%%%%%%%%%%%
%Bibliography
\bibliography{TLN_References}

\bibliographystyle{./utphys1}
%%%%%%%%%%%%%%%%%%%%%%%%%%%%%%%%%%%%%%%%%%%%%%%%%%%%%%%%%%%%%%%%%%%%%%%%%%%%%%%%%%%%%%%%%%%%%%%%%%%
%%%%%%%%%%%%%%%%%%%%%%%%%%%%%%%%%%%%%%%%%%%%%%%%%%%%%%%%%%%%%%%%%%%%%%%%%%%%%%%%%%%%%%%%%%%%%%%%%%%
%%%%%%%%%%%%%%%%%%%%%%%%%%%%%%%%%%%%%%%%%%%%%%%%%%%%%%%%%%%%%%%%%%%%%%%%%%%%%%%%%%%%%%%%%%%%%%%%%%%
\end{document}